\documentclass[11pt]{article}

\usepackage{color,graphicx}
\usepackage{young}
\usepackage[vcentermath]{youngtab}
\usepackage{amsmath,amssymb,graphicx}
\usepackage{hyperref}
\usepackage{float}
\definecolor{darkred}{rgb}{0.65,0.15,0}
\hypersetup{pdfborder={0 0 0},colorlinks=true,urlcolor=darkred,citecolor=blue,linkcolor=darkred,linktocpage=true}

\usepackage{cite}
\usepackage{amsmath}
\usepackage{amsfonts}
\usepackage{amssymb}
\usepackage{graphicx}%
\usepackage{amsthm}
\usepackage{mathrsfs}
\usepackage[T1]{fontenc}
\usepackage{enumerate}
\usepackage{color}
\usepackage{amsmath,
			amssymb,
			amsthm,
			amscd,
			amsfonts,
			amstext,
			indentfirst,
			color,
			slashed,
                   cite}

\usepackage{eucal,dsfont,enumitem}
\usepackage[all]{xy}
\usepackage[active]{srcltx}
\setlist{nolistsep}

\usepackage[top=0.7in,
			bottom=1in,
			left=1in,
			right=1in,
			includehead]
			{geometry}
			
\usepackage{tikz}
\usetikzlibrary{positioning,arrows}
\tikzset{
			>=stealth',
			format/.style={
					  rectangle,
					  rounded corners,
                      minimum size=0.5cm,
                      draw=black,very thick,
                      font=\fontsize{8}{10}\selectfont},
			arrow/.style={
					  ->,
					  thick,
					  shorten <=2pt,
					  shorten >=2pt,}
		}

\numberwithin{equation}{section}

\usepackage{setspace}
\def\4diml{four-dimensional}

\def\-1{^{-1}}

\newcommand{\A}{\mathscr{A}}

\newcommand{\C}{\mathscr{C}}
\newcommand{\M}{\mathscr{M}}
\newcommand{\D}{\mathscr{D}}
\newcommand{\G}{\mathscr{G}}

\newcommand{\tG}{\widetilde{\mathscr{G}}}

\makeatletter

\@addtoreset{equation}{section}
\makeatother

\begin{document}

\thispagestyle{empty}

\vspace{5mm}

\begin{center}
{\LARGE \bf  A hierarchy of WZW models related to super \\[2mm] Poisson-Lie T-duality}

\vspace{15mm}
\normalsize
{\large  Ali Eghbali\footnote{eghbali978@gmail.com},  Adel Rezaei-Aghdam\footnote{rezaei-a@azaruniv.ac.ir}
}

\vspace{2mm}
{\small \em Department of Physics, Faculty of Basic Sciences,\\
Azarbaijan Shahid Madani University, 53714-161, Tabriz, Iran}\\

\vspace{7mm}


\vspace{6mm}

\begin{tabular}{p{12cm}}
{\small
Motivated by super Poisson-Lie (PL) symmetry of the Wess-Zumino-Witten (WZW) model based on the $(C^3+A)$ Lie supergroup of our previous
work [A. Eghbali {\it et al.} JHEP 07 (2013) 134], we first obtain and classify all  Drinfeld superdoubles (DSDs) generated by the
Lie superbialgebra structures on the $({\C}^3+ {\A})$ Lie superalgebra as a theorem.
Then, introducing a general formulation we find the conditions under which a two-dimensional $\sigma$-model
may be equivalent to a WZW model.
With the help of this formulation and starting the super PL symmetric $(C^3+A)$ WZW model, we get a hierarchy of WZW models related to super PL T-duality, in such a way that it is different
from the super PL T-plurality, because the DSDs are, in this process, non-isomorphic.
The most interesting indication of this work is that the $(C^3+A)$ WZW model
does remain invariant under the
super PL T-duality transformation, that is, the model is super PL self-dual.
}
\end{tabular}
\vspace{-1mm}
\end{center}

{Keywords:}  Sigma Models, Conformal and W Symmetry, Bosonic Strings, Conformal Field\\
~~~~~ Models in String Theory, String Duality

\setcounter{page}{1}
\newpage
\tableofcontents

 \vspace{5mm}

\section{\label{Sec.I} Introduction}

Over the past four decades, two-dimensional $\sigma$-models with a supermanifold as target space
have emerged in an immense variety of systems and their study has become increasingly
relevant for some of the challenging problems of modern physics, ranging from e.g.
the condensed matter physics, superstring models and the quantum Hall effect to the AdS/CFT correspondence in string theory.
In condensed matter physics, it has been shown that supermanifolds arise mostly in studying geometrical
problems such as percolation and polymers \cite{G.Parisi}, and also in non-interacting disordered
systems \cite{{Efetov},{Zirnbauer}}.
In string theory, supermanifold target spaces were argued to arise as mirrors of rigid Calabi-Yau (CY) manifolds, i.e.
of CY spaces without complex moduli. The mirror image
of such spaces has no K$\ddot{a}$hler moduli and hence it cannot be a usual CY manifold.
Sethi argued that the dual of a rigid CY is instead given by a CY supermanifold \cite{Sethi}.
In superstring theory, the $\sigma$-models constructed on supermanifolds naturally appear in
the quantization of superstring theory with Ramond-Ramond (RR) backgrounds.
These models have obvious target space supersymmetry and have no worldsheet spinors.

On the other hand, as shown in \cite{{Schomerus1},{Schomerus2},{Schomerus3}}, the WZW models on Lie supergroups are related to local
logarithmic conformal field theories.
The first attempt in this direction dates back to about four decades ago \cite{Henneaux}, where the flat space Green-Schwarz
superstring action was reproduced as a WZW type $\sigma$-model on the coset superspace
(D=10 Poincare supergroup/$SO(9,1))$.
Then, this work was extended to the curved background and
shown that \cite{Metsaev} type IIB superstring on $AdS_5 \times S^5$ can be constructed from a $\sigma$-model
on the coset superspace $SU (2,2|4)/SO(4,1)\times SO(5)$.
In addition, it was shown that the $\sigma$-model on the $PSU(1, 1|2)$ Lie supergroup can
be used to quantize superstring theory on the $AdS_3 \times S^3$ background with RR flux
\cite{{Berkovits1},{Bershadsky}}.
In \cite{Berkovits2}, it was given a formulation of string theory on an interesting four-dimensional background
involving RR fields, in such a way that the $\sigma$-model defined on the coset supermanifold
$PSU(1, 1|2)/U(1) \times U(1)$ was used for quantizing superstring theory on the $AdS_2  \times S^2$
background with RR flux.
In most cases, the models of interest are more complicated than the WZW models constructing on Lie supergroups.
Nevertheless, even WZW models on Lie supergroups are not well understood.
In fact, it is notoriously difficult to analyze the WZW model based on Lie supergroups, even for the
simplest case of $GL(1|1)$ \cite{Saleur} (see also \cite{{Schomerus1},{Schomerus2},{Schomerus3}}).

The duality is an important concept in string theory that seems to be related to integrability. It connects different theories in many perspectives.
As shown, the integrability \cite{{Sfetsos1},{Klimcik1},{Delduc1}} and PL T-duality, introduced in \cite{{Klimcik2},{Klimcik3}},
have in particularly revealed to be strictly connected,
the latter being a generalization of the string T-duality of $\sigma$-models with toric-compactified
backgrounds.
PL T-duality, which is a genuine generalization of T-duality, does not require isomeric symmetries of the target space at all.
Specifically, symmetry under PL T-duality
transformations is based on the concept of PL dual groups and Drinfeld doubles \cite{Drinfeld}.
Generalization of T-duality to Lie supergroups \cite{ER2} (see also \cite{falk}) and also supermanifolds \cite{ER5} has been explored in the context of
PL T-duality, originally introduced as an extension of standard bosonic dualization \cite{{Klimcik2},{Klimcik3}}.

For the sake of clarity, let us comment on some related previous works.
As the first example of the PL symmetry in the WZW models, it was shown that \cite{Klimcik4}
the PL T-duality relates the WZW model based on the SL$(2,\mathbb{R})$ Lie group to a $\sigma$-model defined on the SL$(2,\mathbb{R})$ group space.
Then, we showed that \cite{eghbali11} the WZW model on the Heisenberg Lie group has the PL symmetry
only when its dual pair is the ${ A}_2 \otimes 2{ A}_1$ Lie group (see also \cite{{EMR13},{Exact}}).
Regarding the super PL symmetry of the WZW models based on Lie supergroups, we first showed that \cite{ER7}
the super PL duality relates the $GL(1|1)$ WZW model to a $\sigma$-model defined on the $GL(1|1)$ Lie
supergroup when the dual Lie supergroup is isomorphic to the ${{B} \otimes {A } \otimes {A}_{1,1}}$. Then, in Ref. \cite{ER8} we showed that the WZW
model on the $(C^3+A)$ Lie supergroup has also super PL symmetry
when the dual Lie supergroup is considered to be isomorphic to the ${C^3  \otimes A_{1,1}}$.
Most interesting, we showed that the dual $\sigma$-model on the ${C^3  \otimes A_{1,1}}$
can be equivalent to a WZW model defined on a Lie supergroup which is isomorphic to the $(C^3+A)$ \cite{ER8}.
There, we announced that this process can be continued. Following \cite{ER8}, in the present work, we shall obtain
a hierarchy of WZW models related to the super PL T-duality in such a way that it is different
from super PL T-plurality \cite{E.nucl2020}. As shown in \cite{sakatani1},
PL T-plurality \cite{vonUnge} can construct a chain of supergravity solutions from a PL symmetric solution.
There, it is pointed out that the target space of a WZW model can be used as a solution to generate a chain of solutions via the PL T-plurality.
Recently, super non-Abelian T-duality of $\sigma$-models on group
manifolds with the emphasis on the T-dualization of $\sigma$-models whose target spaces are supermanifolds have received
considerable attention \cite{Bielli1} (see also \cite{Bielli2}), in such a way that it has been performed super non-Abelian T-dualization
of the principal chiral model based on the OSP$(1|2)$ Lie supergroup.
In addition, lately, by studying the obtained geometries by performing super non-Abelian T-duality
of the principal chiral model on the OSP$(1|2)$, it has been shown that \cite{Bielli3} the original model
represents an appropriate three-dimensional supergravity background, interpretable as the superspace version of $AdS_3$, while the
T-dual model fails solving the three-dimensional supergravity torsion constraints.

As a spin off from this progress, it would be interesting to get a hierarchy of WZW models related to super PL T-duality.
In this work, we first find that which of the Manin supertriples
with the first sub-superalgebra $({\C}^3 + {\A})$ represent decomposition of the same (or more precisely isomorphic) DSD.
In Ref. \cite{E.nucl2020}, it has been obtained the DSDs with the first sub-superalgebra $({\C}^3 + {\A})$  which are only isomorphic to the semi-Abelian DSD
$ (({\C}^3 +{\A}) , {\cal I}_{_{(2|2)}})$. Then, using the super PL T-plurality formulation presented in \cite{E.nucl2020} and starting from decompositions of the
$ (({\C}^3 +{\A}) , {\cal I}_{_{(2|2)}})$, it has been found the conformal duality chain of $2+1$-dimensional
cosmological string backgrounds coupling with two fermionic fields (see also \cite{EH-SR}).
The main goal of the present work is to show, using the example of the $(C^3+A)$ WZW model \cite{ER8},
that there is a hierarchy of WZW models related to super PL T-duality.
Most importantly, we show that the WZW model on the $(C^3+A)$  does remain invariant under the
super PL T-duality transformation, that is, the model is super PL self-dual\footnote{The classical self-duality of WZW model can be discussed for a general Lie group $R$. Recently, the extension of the classical
self-duality to a quantum one has been performed by Sakatani {\it et al.} \cite{sakatani2}, in such a way that in order to
discuss the quantum self-duality, they have restricted the Lie group $R$
to be $SU(2)$.}.
It is worth noting that for the bosonic case, we have checked all examples related to WZW models
based on the Lie groups in three dimensions (the $SL(2 , \mathbb{R})$ and $SO(3)$ WZW models) and also
four-dimensional WZW models on the Lie groups  $H_4$ \cite{Exact}, $GL(2 , \mathbb{R})$ \cite{eghbali11}
and Nappi-Witten \cite{witten}. None of them had the conditions related to the example of the $(C^3+A)$ WZW model.
In fact, a hierarchy of the WZW models related to the super PL T-duality is proposed for the first time here.
According to these results we think that, in general, it is not possible to prove the mechanism shown (hierarchy of WZW models) in this article for any arbitrary DSD.

The paper is organized as follows. In section \ref{Sec.II}, we determine Manin supertriples
with the first sub-superalgebra $({\C}^3+ {\A})$ and show that they can be belonged to the 24 classes of the non-isomorphic DSDs.
In section \ref{Sec.III}, we briefly review the construction of the super PL T-dualizable
$\sigma$-models on supermanifolds. In section \ref{Sec.IV},
we find the conditions under which a two-dimensional $\sigma$-model may be equivalent to a WZW model.
Section \ref{Sec.V} contains the original results of the work:  using the general formulation introduced in section \ref{Sec.IV} and then starting the super PL symmetric $(C^3+A)$ WZW model, we get a hierarchy of WZW models related to super PL T-duality.
Some concluding remarks are given in the last section.
At the end of the article we added Appendix A devoted to carry
out the DSD isomorphisms among the corresponding Manin supertriples.

\section{\label{Sec.II} DSDs generated by Manin supertriples
with first sub-superalgebra $({\C}^3 + {\A})$}

\subsection{\label{Sec.II.1} A review of DSDs on the algebraic level}
Let us start with the definition of a Lie superalgebra. A Lie superalgebra ${\G}$ is firstly a $Z_2$-graded vector space,
thus admitting the decomposition ${\G} ={\G}_{{_B}} \oplus {\G}_{_F}$
(we refer to ${\G}_{{_B}}$ and ${\G}_{{_F}}$ as the even and odd subspaces of ${\G}$, respectively\footnote{In theoretical physics,
the even elements are sometimes called Bose elements or bosonic, and the odd elements Fermi elements or fermionic.}), secondly there is defined on ${\G}$
a particular kind of binary operation, i.e., a mapping $ {\G} \otimes {\G} \rightarrow {\G}$,
denoted by $[. , .]$ satisfying the requirements of super antisymmetry and super Jacobi identity \cite{{N.A},{J.z},{ER1}}.
One says that the Lie superalgebra ${\G}$ is of the superdimension $(m | n)$ iff $dim\;{\G}_{{_B}} = m$
and $dim~{\G}_{{_F}} = n$. It may be finite or infinite.
To evade the problems with definitions of supergroups we shall define the
DSDs only on the algebraic level.

The DSD $D$ is defined as a Lie supergroup whose Lie superalgebra, ${\D}$, equipped
by a supersymmetric ad-invariant non-degenerate bilinear form
$<.~,~.>$ can be decomposed into
a pair of maximally isotropic sub-superalgebras ${\G}$ and ${\tilde {\G}}$,
and ${\D}$ as a vector superspace is the direct sum of ${\G}$ and ${\tilde {\G}}$.
This ordered triple of Lie superalgebras $({\D}, {\G}, {\tilde {\G}})$ is called {\it Manin supertriple}.
Moreover, the pair of $({\G} , {\tilde {\G}})$ is called {\it Lie superbialgebra}\footnote{Notice that we consider
${\tilde {\G}}$ as a dual ${{\G}^\ast}$ to $\G$
and  use the Lie superbracket in ${\tilde {\G}}$ to define the Lie super cobracket in $\G$ \cite{ER1}.} \cite{{N.A},{J.z},{ER1}}.
Note that the dimensions of the sub-superalgebras are equal and
the bases ${T_{_a}}$ and ${\tilde T}^{^a}, a= 1,...,dim~{\G}$ in the sub-superalgebras can be chosen
so that\footnote{Similar to our previous works, here we consider the standard basis for the supervector spaces so that in writing the basis
as a column matrix, we first present the bosonic bases, then the fermionic ones (for further details refer to Ref. \cite{D}).
We note that $|a|$ denotes the parity of  $a$ where $|a|=0$ for the bosonic coordinates and $|a|=1$ for the
fermionic ones.  Throughout the article we use Dewitt's notation \cite{D} such that
we identify the grading indices by the same indices in the power of $(-1)$, that is, we use $(-1)^a$ instead of  $(-1)^{|a|}$.}
\begin{eqnarray}\label{2.1}
<{T_{_a}} , {T_{_b}}> =0,~~~<{\tilde T}^{^a}  , {\tilde T}^{^b}> =0,~~~~
{{\delta}^{^b}}_{a} =   <{\tilde T}^{^b} , {T_{_a}} > = (-1)^{^{|a||b|}}  <{T_{_a}} , {\tilde T}^{^b}>.
\end{eqnarray}
Due to the ad-invariance of bilinear form
$<. , .>$ the algebraic
structure of ${\D}$ is determined by the structure of the maximally
isotropic sub-superalgebras because in the bases ${T_{_a}}$ and ${\tilde T}^{^a}$ the Lie superbrackets
are given by
\begin{eqnarray}\label{2.2}
[{T_{_a}} , {T_{_b}}] &=& {f^c}_{_{ab}} ~{T_{_c}},~~~~~~~
[{\tilde T}^{^a} , {\tilde T}^{^b}] = {{\tilde f}^{ab}}_{\;   \; c} ~{\tilde T}^{^c},\nonumber\\
{[{T_{_a}} , {\tilde T}^{^b}]} &=& (-1)^{^b} {{{\tilde f}^{bc}}}_{\; \;a}~ {T_{_c}} + (-1)^{^a} {f^b}_{_{ca}} ~{\tilde T}^{^c}.
\end{eqnarray}
The Lie superalgebra structure defined by relation \eqref{2.2} is called the DSD ${\D}$.
It is worth mentioning that the super Jacobi identity of Lie superalgebra ${\D}$ imposes the following relation over the structure
constants of Lie superalgebras  ${\G}$ and ${\tilde {\G}}$ \cite{ER1}
\begin{eqnarray}\label{2.3}
{f^d}_{bc}{\tilde{f}^{ae}}_{\; \; \; \; d}=
{f^a}_{dc}{\tilde{f}^{de}}_{\; \; \; \; \; b} +
{f^e}_{bd}{\tilde{f}^{ad}}_{\; \; \; \; \; c}+ (-1)^{be}
{f^a}_{bd}{\tilde{f}^{de}}_{\; \; \; \; \; c}+ (-1)^{ac}
{f^e}_{dc}{\tilde{f}^{ad}}_{\; \; \; \; \; b}.
\end{eqnarray}

\subsection{\label{Sec.II.2} Manin supertriples with first sub-superalgebra $({\C}^3 + {\A})$}

Since we are interested on the classification of DSDs generated by Manin supertriples
with first sub-superalgebra $({\C}^3 + {\A})$, as the first step we define the $({\C}^3 +{\A})$ Lie superalgebra.
The $({\C}^3 +{\A})$ is a four-dimensional Lie superalgebra of the type $(2 | 2)$ which
is spanned by the set of generators $\{T_{_1}, T_{_2}; T_{_3}, T_{_4}\}$ with gradings;
grade$(T_{_1})$=grade$(T_{_2})=0$ and grade$(T_{_3})$=grade$(T_{_4})=1$ \footnote{From now on we
consider $(T_{_1}, T_{_2})$ and $(T_{_3}, T_{_4})$ as bosonic and fermionic generators, respectively.}. These generators
fulfill the following  non-zero (anti)commutation rules  \cite{B}:
\begin{eqnarray}\label{2.4}
[T_{_1} , T_{_4}]=T_{_3},~~~~~~~\{T_{_4} , T_{_4}\}=T_{_2}.
\end{eqnarray}
It is also interesting to note that the $({\C}^3 + {\A})$ is itself a
two-dimensional Lie superbialgebras, i.e., it is isomorphic to the $(2|2)$-dimensional DSD $(({\A}_{_{1,1}}+{\A}) , {\cal I}_{_{(1|1)}})$ \cite{ER6}.
Note that the classification of four-dimensional Lie superalgebras of the type $(2 | 2)$
was, for the first time, performed by Backhouse in \cite{B} (see also \cite{{ER3},{ER6}} for the classification of
decomposable Lie superalgebras of the types $(1|2)$, $(2|1)$ and $(2|2)$).

In Ref. \cite{ER14} we performed a complete classification of the $({\C}^3 +{\A})$ Lie superbialgebra
and obtained 31 families of inequivalent Lie superbialgebra
structures on $({\C}^3 + {\A})$ whose representatives have classified
in Table 1.
Based on Table 1, the label of each Manin supertriple, e.g. $\big(({\C}^3 + {\A}) , ({\C}^3 + {\A})_k^{\epsilon= \pm 1}\big)$,
indicates the structure of the first sub-superalgebra $\G$, e.g.  $({\C}^3 + {\A})$, the
structure of the second sub-superalgebra $\tG$, e.g.  $({\C}^3 + {\A})_k^{\epsilon = \pm 1}$; roman numbers $i$, $ii$ etc. (if
present) distinguish between several possible pairings $<. , .>$ of the sub-superalgebras $\G$ and $\tG$,
and the parameter $k$  indicates the Manin supertriples differing in the rescaling of  $<. , .>$.
As we use DeWitt notation here, in the standard basis the structure constant
${f^{^B}}_{_{FF}}$ (anticommutator of fermion-fermion) must be pure imaginary. Despite of this  we have omitted
the coefficient $i = \sqrt{-1}$ from all anticommutation relations in Table 1.

\vspace{3mm}

\hspace{0.10cm}{\footnotesize  Table 1.} {\small  Dual Lie
superalgebras to the $({\C}^3+{\A})$ Lie superalgebra. }\\
\begin{tabular}{l l l   p{15mm} }
\hline\hline

{\footnotesize $ \tilde {\G}$ }& {\footnotesize Non-zero
(anti)commutation relations}&{\footnotesize Comments}  \smallskip\\
\hline
\smallskip

\vspace{0.5mm}

{\scriptsize ${\cal I}_{(2 | 2)}$}& {\scriptsize All of the (anti)commutation relations are zero.} \\

\vspace{1mm}

{\scriptsize ${\C}_{p=1}^{1,\epsilon} \oplus {\A}$}&{\scriptsize
$[{\tilde T}^{^1},{\tilde T}^{^2}]= \epsilon {\tilde T}^{^1},\;\;\; [{\tilde T}^{^2},{\tilde T}^{^3}]= -\epsilon {\tilde T}^{^3}$}&\\

\vspace{1mm}

{\scriptsize ${\C}_{p=-1}^{1,\epsilon} \oplus {\A}$}&{\scriptsize
$[{\tilde T}^{^1},{\tilde T}^{^2}]= \epsilon {\tilde T}^{^1},\;\;\; [{\tilde T}^{^2},{\tilde T}^{^4}]= \epsilon {\tilde T}^{^4}$}&\\

\vspace{1mm}

{\scriptsize ${\C}_{p=1}^{2,\epsilon} \oplus {\A}_{1,1}$}&{\scriptsize
$[{\tilde T}^{^2},{\tilde T}^{^3}]= \epsilon {\tilde T}^{^3},\;\;\; [{\tilde T}^{^2},{\tilde T}^{^4}]= \epsilon {\tilde T}^{^4}$}&\\

\vspace{1mm}

{\scriptsize ${\C}^{3} \oplus {\A}_{1,1}.i$}&{\scriptsize
$[{\tilde T}^{^2},{\tilde T}^{^3}]=  {\tilde T}^{^4}$}&\\

\vspace{1mm}

{\scriptsize ${\C}^{4,\epsilon} \oplus {\A}_{1,1}$}&{\scriptsize
$[{\tilde T}^{^2} , {\tilde T}^{^3}]=  {\tilde T}^{^4},\;\;\; [{\tilde T}^{^2},{\tilde T}^{^3}]= \epsilon {\tilde T}^{^3},\;\;\;
[{\tilde T}^{^2} ,{\tilde T}^{^4}]= \epsilon {\tilde T}^{^4}$}&\\

\vspace{1mm}

{\scriptsize ${\D}_{p,p-1}^{1,\epsilon}$}&{\scriptsize
$[{\tilde T}^{^1},{\tilde T}^{^2}]= -\epsilon {\tilde T}^{^1},\;\;\; [{\tilde T}^{^2},{\tilde T}^{^3}]= \epsilon p {\tilde T}^{^3},\;\;\;
[{\tilde T}^{^2},{\tilde T}^{^4}]= \epsilon(p-1) {\tilde T}^{^4}$}& {\scriptsize $p \neq 0,1$}\\

\vspace{1mm}

{\scriptsize $\big({\D}_{\frac{1}{2},\frac{1}{2}}^{7,\epsilon}\big)^2$}&{\scriptsize
$[{\tilde T}^{^1},{\tilde T}^{^2}]= -2\epsilon {\tilde T}^{^1},\;\;\; [{\tilde T}^{^2},{\tilde T}^{^3}]= \epsilon  {\tilde T}^{^3},\;\;\;
[{\tilde T}^{^2},{\tilde T}^{^4}]= \epsilon {\tilde T}^{^4},\;\;\;\{{\tilde T}^{^3},{\tilde T}^{^4}\}= 2 \epsilon {\tilde T}^{^1} $}& \\

\vspace{-1mm}

{\scriptsize $\big({\D}_{1-p,p}^{7,\epsilon}\big).i$}&{\scriptsize
$[\tilde T^1,\tilde T^2]= -\epsilon \tilde T^1,\;\;\; [\tilde T^2,\tilde T^3]= \epsilon p  \tilde T^3,\;\;\;
[\tilde T^2,\tilde T^4]= -\epsilon (p-1) \tilde T^4,$}&\\

&{\scriptsize $ \{\tilde T^3,\tilde T^4\}= -2 \epsilon (p-1) \tilde T^1 $}& {\scriptsize $ p<\frac{1}{2},~p \neq 0$}\\

\vspace{-1mm}

{\scriptsize $\big({\D}_{1-p,p}^{7,\epsilon}\big).ii$}&{\scriptsize
$[\tilde T^1,\tilde T^2]= -\epsilon \tilde T^1,\;\;\; [\tilde T^2,\tilde T^3]= -\epsilon (p-1) \tilde T^3,\;\;\;
[\tilde T^2,\tilde T^4]= \epsilon p \tilde T^4,$}&\\

&{\scriptsize $ \{\tilde T^3,\tilde T^4\}= 2 \epsilon p \tilde T^1 $}& {\scriptsize $ p<\frac{1}{2},~p \neq 0$}\\

\vspace{-1mm}

{\scriptsize $\big({\D}_{0}^{10,\epsilon}\big)^1$}&{\scriptsize
$[\tilde T^1,\tilde T^2]= -\epsilon \tilde T^1,\;\;\; [\tilde T^1,\tilde T^3]= -\frac{\epsilon}{4}  \tilde T^4,\;\;\;
[\tilde T^2,\tilde T^4]= \epsilon \tilde T^4, ~~~\{\tilde T^3,\tilde T^3\}=  \epsilon \tilde T^2,$}&\\

&{\scriptsize $\{\tilde T^3,\tilde T^4\}= 2\epsilon \tilde T^1$}&\\

\vspace{-1mm}

{\scriptsize $\big({\D}_{0}^{10,\epsilon}\big)^2$}&{\scriptsize
$[\tilde T^1,\tilde T^2]= -\epsilon \tilde T^1,\;\;\; [\tilde T^1,\tilde T^3]= \frac{\epsilon}{4}  \tilde T^4,\;\;\;
[\tilde T^2,\tilde T^4]= \epsilon \tilde T^4, ~~~\{\tilde T^3,\tilde T^3\}= - \epsilon \tilde T^2,$}&\\

&{\scriptsize $\{\tilde T^3,\tilde T^4\}= 2\epsilon \tilde X^1$}&\\

\vspace{1mm}

{\scriptsize $\big(2{\A}_{1,1} + 2 {\A} \big)^0.i$}&{\scriptsize
$\{{\tilde T}^{^3},{\tilde T}^{^3}\}=  {\tilde T}^{^1}$}&\\

\vspace{1mm}

{\scriptsize $\big({\C}_{1}^1 + {\A}\big)^\epsilon$}&{\scriptsize
$[{\tilde T}^{^1} ,{\tilde T}^{^2}]= -\frac{\epsilon}{2} {\tilde T}^{^1},\;\;\;[{\tilde T}^{^2},{\tilde T}^{^4}]= \frac{\epsilon}{2} {\tilde T}^{^4}, ~~~
\{{\tilde T}^{^3},{\tilde T}^{^4}\}= \epsilon {\tilde T}^{^1}$}&\\

\vspace{1mm}

{\scriptsize $\big({\C}_{-1}^2 + {\A}\big)^\epsilon$}&{\scriptsize
$[{\tilde T}^{^2},{\tilde T}^{^3}]= {\epsilon} {\tilde T}^{^3},\;\;\;[{\tilde T}^{^2},{\tilde T}^{^4}]=-{\epsilon} {\tilde T}^{^4}, ~~~
\{{\tilde T}^{^3},{\tilde T}^{^4}\}=-2 \epsilon {\tilde T}^{^1}$}&\\

\vspace{1mm}

{\scriptsize $\big({\C}^3 + {\A}\big)^\epsilon$}&{\scriptsize
$[{\tilde T}^{^1} ,{\tilde T}^{^3}]= -\frac{\epsilon}{4}{\tilde T}^{^4},\;\;\;
\{{\tilde T}^{^3} ,{\tilde T}^{^3}\}=\epsilon {\tilde T}^{^2}$}&\\

\vspace{1mm}

{\scriptsize $\big({\C}^3 + {\A}\big)_k^\epsilon$}&{\scriptsize
$[{\tilde T}^{^2} ,{\tilde T}^{^3}]= {\epsilon} {\tilde T}^{^4},\;\;\;
\{{\tilde T}^{^3},{\tilde T}^{^3}\}= k {\tilde T}^{^1}$}&{\scriptsize $k>0$}\smallskip\\

\hline\hline
\end{tabular}\\
{\small ${^{\ast}}{\epsilon}  = \pm 1;$~grade$({\tilde T}^{^1})=$grade$({\tilde T}^{^2})=0$,~grade$({\tilde T}^{^3})=$grade$({\tilde T}^{^4})=1$.}
\\

In the following, we consider how many of $(4|4)$-dimensional Lie superalgebras $\D$ of DSD
D where related to the $({\C}^3+{\A})$ Lie
superbialgebras of Tables 1 are isomorphic. To this end, we consider two DSDs isomorphic
if they have isomorphic algebraic structure and there is an isomorphism matrix transforming
one ad-invariant bilinear form to the other.
Let $\D$ and $\D'$ be two DSDs with the bases $X_{_A}=(T_{_a} , {\tilde T}^{^b})$ and $X'_{_A}=({T'}_{_a} , {\tilde {T'}}^{^b})$, respectively, so that
the bases satisfy equations \eqref{2.1} and \eqref{2.2} with the structure constants ${F^{^C}}_{_{AB}}$ and ${F'^{^C}}_{_{AB}}$ of the
doubles $\D$ and ${\D'}$. By definition, the DSDs  $\D$ and ${\D'}$ are isomorphic to each other iff
if there is a linear bijection that transforms grading, bilinear form and algebraic structure of $\D$ into those of ${\D'}$.
This means that in the standard basis there is a block diagonal isomorphism matrix $\mathbb{C}=$diag$(\mathbb{C}_{_B} , \mathbb{C}_{_F})$
between $\D$ and ${\D'}$ ($\D \overset{\mathbb{C}}{\longrightarrow} \D'$) such that ${\D}_{_B} \longrightarrow {\D'}_{_B}$ and ${\D}_{_F} \longrightarrow {\D'}_{_F}$.
Accordingly, the linear map given by
\begin{eqnarray}\label{2.5}
X'_{_A} = (-1)^{^B}~ {\mathbb{C}_{_{A}}}^{^B}~ X_{_B},
\end{eqnarray}
transforms the Lie multiplication of $\D$ into that of ${\D'}$ and preserves
the canonical form of the bilinear form $<. , .>$. This is equivalent to\footnote{``st'' denotes supertransposition.}
\begin{eqnarray}
(-1)^{^{AB+ED}}~ \mathbb{C}~ {\cal Y}^{^E}~ \mathbb{C}^{^{st}} &=& {\cal Y'}^{^C}~ {{\mathbb{C}}_{_C}}^{^E},\label{2.6}\\
(-1)^{^{A}}~ \mathbb{C}~ \mathbb{B}~ \mathbb{C}^{^{st}} &=& \mathbb{B},\label{2.7}
\end{eqnarray}
where $({\cal Y}^{^E})_{_{AB}} = - {F^{^E}}_{_{AB}}$ and $({\cal Y'}^{^E})_{_{AB}} = - {F'^{^E}}_{_{AB}}$ are the matrix representations
of the Lie superalgebras $\D$ and ${\D'}$, respectively. In equation \eqref{2.6}, the indices $A$ and
$B$ correspond to the row and column of matrix ${\cal Y}^{^E}$, respectively, and $D$ denotes the column of
matrix $\mathbb{C}^{^{st}}$. In equation \eqref{2.7}, the index $A$ denotes the row of
matrix $\mathbb{B}$ so that $\mathbb{B}$ is the block matrix of
the bilinear form $<. , .>$.
In the standard basis given by \eqref{2.1} it is then read
{\small \begin{eqnarray}\label{2.8}
\mathbb{B}_{_{AB}}=<X_{_A} , X_{_B}>=\left( \begin{tabular}{cc|cc}
                 0 & ${\bf 1}_{_m}$ & 0 & 0 \\
                 ${\bf 1}_{_m}$ & 0 & 0 & 0 \\ \hline
                 0 & 0 & 0 & -${\bf 1}_{_n}$ \\
                0 & 0 & ${\bf 1}_{_n}$ & 0 \\
                 \end{tabular} \right),
\end{eqnarray}}
where ${\bf 1}_{_{m(n)}}$ is the identity matrix of dimension $m(n)$. It is obvious that the superdimension
of these DSDs is $(2m|2n)$.

As was mentioned earlier, all Manin supertriples with the first sub-superalgebra $({\C}^3 + {\A})$ have listed in Table 1.
We are going to classify the corresponding DSDs by looking for the classes of
Manin supertriples that are isomorphic as DSDs. In this way, we apply the method mentioned above
to determine non-isomorphic DSDs as well as the transformation between Manin supertriples that are isomorphic.
The DSD isomorphisms among the
corresponding Manin supertriples are given in Appendix A.
The final distinction between non-isomorphic DSDs and their decomposition into Manin supertriples
provides the following theorem.
\\\\
{\bf Theorem 1:}~{\it Any DSD generated by Manin supertriples
with the first sub-superalgebra ${\cal G}\equiv({\C}^3+ {\A})$ belongs
just to one of the following 24 classes and allows decomposition into non-isomorphic
Manin supertriples listed in the class and their duals}\\

 \begin{tabular}{l l l l l p{0.15mm} }

  \vspace{1mm}
{\footnotesize  $DSD 1:$ } & {\footnotesize  $\big({\cal G} , {\cal {I}}_{_{(2|2)}}\big) \cong
\big({\cal G} , {{\C}}^3 \oplus {{\A}_{_{1,1}}}.i\big) \cong  \big({\cal G} , {({{\C}}^3 + {\A})_{_{k=4}}}^{\hspace{-5mm}\epsilon=-1}\big)$ } \\

\vspace{1mm}
{\footnotesize  $DSD 2:$ } & {\footnotesize  $\big({\cal G} , {{\C}_{_{p=1}}}^{\hspace{-4mm}1,\epsilon=1} \oplus {{\A}}\big)$ } \\
 \vspace{1mm}
{\footnotesize  $DSD 3:$ } & {\footnotesize  $\big({\cal G} , {{\C}_{_{p=1}}}^{\hspace{-4mm}1,\epsilon=-1} \oplus {{\A}}\big)$  } \\

\vspace{1mm}
{\footnotesize  $DSD 4:$ } & {\footnotesize  $\big({\cal G} , {{\C}_{_{p=-1}}}^{\hspace{-6mm}1,\epsilon=1} \oplus {{\A}}\big)\cong
\big({\cal G} , {{\C}_{_{p=-1}}}^{\hspace{-6mm}1,\epsilon=-1} \oplus {{\A}}\big)$ } \\
  \vspace{1mm}
{\footnotesize  $DSD 5:$ } & {\footnotesize  $\big({\cal G} , {{\C}_{_{p=1}}}^{\hspace{-4mm}2,\epsilon=1} \oplus {{\A}_{_{1,1}}}\big)\cong
\big({\cal G} , {{\C}_{_{p=1}}}^{\hspace{-4mm}2,\epsilon=-1} \oplus {{\A}_{_{1,1}}}\big)\cong
\big({\cal G} , {{\C}}^{4,\epsilon=1} \oplus {{\A}_{_{1,1}}}\big)\cong
\big({\cal G} , {{\C}}^{4,\epsilon=-1} \oplus {{\A}_{_{1,1}}}\big)$ } \\
\vspace{1mm}
{\footnotesize   $DSD 6:$ } & {\footnotesize $\big({\cal G} , ({{\D}_{_{\frac{1}{2}, \frac{1}{2}}}}^{\hspace{-4mm}7,\epsilon=1})^2\big)$} \\
\vspace{1mm}
{\footnotesize   $DSD 7:$ } & {\footnotesize  $\big({\cal G} , ({{\D}_{_{\frac{1}{2}, \frac{1}{2}}}}^{\hspace{-4mm}7,\epsilon=-1})^2\big)$ } \\

 \vspace{1mm}
{\footnotesize   ${DSD 8}^{^{p}}$~~: } & {\footnotesize  $\big({\cal G} , ({{\D}_{_{1-p, p}}}^{\hspace{-6mm}7,\epsilon=1}).i\big),~
~~~~~~p\neq0,~ p<\frac{1}{2}$ } \\

\end{tabular}

\newpage
\begin{tabular}{l l  p{0.15mm} }

 \vspace{1mm}
{\footnotesize   ${DSD 9}^{^{p}}$~: } & {\footnotesize  $\big({\cal G} , ({{\D}_{_{1-p, p}}}^{\hspace{-6mm}7,\epsilon=-1}).i\big),~
~~~~p\neq0,~ p<\frac{1}{2}$ } \\

 \vspace{1mm}
{\footnotesize   ${DSD 10}^{^{p}}$~: } & {\footnotesize  $\big({\cal G} , ({{\D}_{_{1-p, p}}}^{\hspace{-6mm}7,\epsilon=1}).ii\big),~
~~~~~p\neq0,~ p<\frac{1}{2}$ } \\

 \vspace{1mm}
{\footnotesize   ${DSD 11}^{^{p}}$~: } & {\footnotesize  $\big({\cal G} , ({{\D}_{_{1-p, p}}}^{\hspace{-6mm}7,\epsilon=-1}).ii\big),~
~~~p\neq0,~ p<\frac{1}{2}$ } \\

 \vspace{1mm}
{\footnotesize   ${DSD 12}$: } & {\footnotesize  $\big({\cal G} , ({{\D}_{_{0}}}^{\hspace{-1mm}10,\epsilon=1})^1\big)$ } \\

\vspace{1mm}
{\footnotesize   ${DSD 13}$: } & {\footnotesize  $\big({\cal G} , ({{\D}_{_{0}}}^{\hspace{-1mm}10,\epsilon=-1})^1\big)$ } \\

\vspace{1mm}
{\footnotesize   ${DSD 14}$: } & {\footnotesize  $\big({\cal G} , ({{\D}_{_{0}}}^{\hspace{-1mm}10,\epsilon=1})^2\big)$ } \\

\vspace{1mm}
{\footnotesize   ${DSD 15}$: } & {\footnotesize  $\big({\cal G} , ({{\D}_{_{0}}}^{\hspace{-1mm}10,\epsilon=-1})^2\big)$ } \\

\vspace{1mm}
{\footnotesize   ${DSD 16}$: } & {\footnotesize  $\big({\cal G} , ({{{\C}_{_{1}}}^{\hspace{-1mm}1} + {\A})}^{\epsilon=1}\big)$ } \\

\vspace{1mm}
{\footnotesize   ${DSD 17}$: } & {\footnotesize  $\big({\cal G} , ({{{\C}_{_{1}}}^{\hspace{-1mm}1} + {\A})}^{\epsilon=-1}\big)$ } \\

\vspace{1mm}
{\footnotesize   ${DSD 18}$: } & {\footnotesize  $\big({\cal G} , ({{{\C}_{_{-1}}}^{\hspace{-2mm}2} + {\A})}^{\epsilon=1}\big)$ } \\

\vspace{1mm}
{\footnotesize   ${DSD 19}$: } & {\footnotesize  $\big({\cal G} , ({{{\C}_{_{-1}}}^{\hspace{-2mm}2} + {\A})}^{\epsilon=-1}\big)$ } \\

\vspace{1mm}
{\footnotesize   ${DSD 20}$: } & {\footnotesize  $\big({\cal G} , ({{{\C}}^{3} + {\A})}^{\epsilon=1}\big)$ } \\

\vspace{1mm}
{\footnotesize   ${DSD 21}$: } & {\footnotesize  $\big({\cal G} , ({{{\C}}^{3} + {\A})}^{\epsilon=-1}\big)$ } \\

\vspace{1mm}
{\footnotesize ${DSD 22}^{^p}:$  } & {\footnotesize  $
\big({\cal G} , { {\D}_{_{p,p-1}}}^{\hspace{-6mm}1,\epsilon=1}\big)\cong
\big({\cal G} , { {\D}_{_{1-p,-p}}}^{\hspace{-7mm}1,\epsilon=1}\big)\cong
\big({\cal G} , { {\D}_{_{p,p-1}}}^{\hspace{-6mm}1,\epsilon=-1}\big),~~~p\neq 0,1$ } \\
\vspace{1mm}
{\footnotesize ${DSD 23}^{^k}:$  } & {\footnotesize  $\big({\cal G} , {(2{\A}_{_{1,1}} + 2{\A})^{^{0}}}.i\big)\cong
\big({\cal G} , {({ {\C}}^3 + {\A})_{_{k-4}}}^{\hspace{-4mm}\epsilon=1}\big)\cong
\big({\cal G} , {({ {\C}}^3 + {\A})_{_{k}}}^{\hspace{-2mm}\epsilon=-1}\big),~~~k>4$ } \\
\vspace{1mm}
{\footnotesize ${DSD 24}^{^k}:$  } & {\footnotesize  $\big({\cal G} , {({{\C}}^3 + {\A})_{_{k}}}^{\hspace{-2mm}\epsilon=-1}\big)\cong
\big({\cal G} , {{({\C}}^3 + {\A})_{_{4-k}}}^{\hspace{-4mm}\epsilon=-1}\big),~~~0<k<4$. } \\
 \end{tabular}
\\

As can be seen from the above theorem, for many DSDs such as $DSD 1,  DSD 4, DSD 5, {DSD 22}^{^p}$, ${DSD 23}^{^k}$ and
${DSD 24}^{^k}$ there are several decompositions into Manin supertriples.
Notice that the structure of $DSD1$ has already been employed in \cite{E.nucl2020}. There, it has shown that this structure gives us the
possibility to apply the super PL T-plurality to $\sigma$-models whose target spaces are
Lie supergroups.


\section{\label{Sec.III} A short review of super PL T-duality}

In this section we briefly review the construction of the super PL T-dualizable
$\sigma$-models on supermanifolds \cite{ER2}.
Let $\M$ be a $(d_{_B}|d_{_F})$-dimensional target supermanifold \footnote{The
superdimension of the supermanifold $\M$ is written as ($\#$ bosons $|$
$\#$ fermions)=$(d_{_B}|d_{_F})$;  because of the invertibility of the
metric $G_{_{MN}}$, $d_{_F}$ must be even \cite{D}.} and consider two-dimensional $\sigma$-model on $\M$ given by action
\begin{eqnarray}\label{4.1}
S &=& \frac{1}{2}\int_{_{\Sigma}}\!d\sigma^+  d\sigma^- ~(-1)^{^M}~ \partial_{_+} \phi^{^M}~{\cal E}_{_{MN}}~\partial_{_-} \phi^{^N},
\end{eqnarray}
where $\partial_{_\pm}$ are the derivatives with respect to the standard lightcone
variables $\sigma^{\pm}=\tau \pm \sigma$ on the worldsheet $\Sigma$.
The functions $\phi^{^M}$  include the bosonic coordinates $x^\mu$ and the fermionic ones $\theta^i$,
and the labels $M, N$ run over $\mu =0,\cdots, d_{_B}- 1$ and $i = 1,\cdots, d_{_F}$.
The tensor field ${\cal E}_{_{MN}}$ can be understood as a sum of the
supersymmetric metric ${G}_{_{MN}}$ and the Kalb-Ramond field ${B}_{_{MN}}$ (or $B$-field that is a super antisymmetric tensor) defining the geometric properties of the supermanifold $\M$.
It  seems  to  be  of  interest to define the line element and $B$-field corresponding to 
action \eqref{4.1} in the coordinate basis. They are, respectively, read 
\begin{eqnarray}
ds^2 &=& (-1)^{^{MN}} ~G_{_{MN}} d\phi^{^M}~d\phi^{^N},\label{4.1.2}\\
B &=& \frac{ (-1)^{^{MN}}}{2} ~B_{_{MN}} ~ d\phi^{^M} \wedge d\phi^{^N}.\label{4.1.3}
\end{eqnarray}
If the Noether's current one-forms corresponding to the right action of the Lie supergroup
$G$ on the target supermanifold $\M$ are not closed and satisfy the super Maurer-Cartan equation \cite{D} on the
extremal surfaces, we say that the $\sigma$-model \eqref{4.1} has the super PL symmetry
with respect to the Lie supergroup $\tilde G$ (the dual Lie supergroup to $G$ with the same superdimension $G$). It is a condition
that is given by the following relation \cite{ER2}
\begin{eqnarray}\label{4.2}
{\cal L}_{_{V_{_a}}}{\cal E}_{_{MN}}\;=\;(-1)^{^{a+Ma+Qc+P}}~{{\tilde f}^{bc}}_{~~a}~ {\cal E}_{_{MQ}}~ {V_{_c}}^{^{~Q}}~{V_{_b}}^{^{~P}}~
{\cal E}_{_{PN}},
\end{eqnarray}
where ${\cal L}_{_{V_{_a}}}$ stands for the Lie derivative corresponding to the left invariant supervector
fields ${V_{_a}}= {V_{_a}}^{^M} {{{\overrightarrow \partial}/{\partial{\phi^{^M}}}
}}$ (defined with left derivative) satisfying $[{V_{_a}} , {V_{_b}}]= {f^c}_{ab} ~{V_{_c}}$, and
${{\tilde f}^{bc}}_{~~a}$ are the structure constants of the dual Lie superalgebra $\tG$ of $\tilde G$.
By direct calculation, one can obtain
\begin{eqnarray}\label{4.3}
{\cal L}_{_{V_{_a}}}{\cal E}_{_{MN}} &=& (-1)^{^{Ma + M+ P}}\;{{\frac{\overrightarrow \partial}{\partial{\phi^{^M}}}
}}\;{V_{_a}}^{^P}\;  {\cal E}_{_{PN}}+{V_{_a}}^{^P}\;{{\frac{\overrightarrow \partial}{\partial{\phi^{^P}}}
}}\;{\cal E}_{_{MN}}\nonumber\\
&&~~~~~~~~~~~~~~~~~~~~~~~~~~~~~~~+(-1)^{^{MN+ MP+Na+N+P}}
\;{{\frac{\overrightarrow \partial}{\partial{\phi^{^N}}}
}}\;{V_{_a}}^{^P}\; {\cal E}_{_{MP} }.
\end{eqnarray}
The integrability condition on the Lie derivative,
$[{\cal L}_{_{V_{_a}}} , {\cal L}_{_{V_{_b}}}]= {\cal L}_{_{[V_{_a} , V_{_b}]}}$,
then implies the mixed super Jacobi identities \eqref{2.3} showing that this construction leads naturally to
the DSD.

Both the original and dual geometries of the super PL
T-dualizable $\sigma$-models are derived from the so-called DSD which is a Lie supergroup.
In the following, we shall consider T-dual $\sigma$-model on
a Lie supergroup $G$. To this end, suppose now that $G$ acts transitively and
freely on the supermanifold $\M$; then the target can be identified with the
supergroup $G$. In this case, the $\sigma$-model having target space in the Lie supergroup $G$
is given by the  following action \cite{ER2}
\begin{eqnarray}\label{4.4}
S\;=\;\frac{1}{2}\int_{_{\Sigma}}\!d\sigma^{+}
d\sigma^{-}\;(-1)^{^{a}} {{R_{_+}}}^{\hspace{-1.5mm} a}\;  {E_{_{ab}}}(g)\;{{R_{_-}}}^{\hspace{-1.5mm} b},
\end{eqnarray}
where ${{R_{_\pm}}}^{\hspace{-1.5mm} a}$ are the components of the right invariant super one-forms
which are defined by means of an element $g: \Sigma \rightarrow G$ in the following formula
\begin{eqnarray}\label{4.5}
 \partial_{_{\pm}} g  ~g^{-1}=(-1)^a ~{R_{_{\pm}}^{a}}~T_{{_a}}=(-1)^a ~\partial_{_\pm} \phi^{^{M}}
\; {{_{_M}} R}^{^a}~T_{{_a}},
\end{eqnarray}
in which $\phi^{^{M}}$ are coordinates of the elements of $G$.
The ${E_{_{ab}}}(g)$  is a certain bilinear form on the ${\G}$ and is defined by
\footnote{Here one must use the superinverse formula introduced in \cite{D}.}
\begin{eqnarray}\label{4.6}
{ E(g)}\;=\;\big({E_{_0}}^{-1} + \Pi(g)\big)^{-1}, ~~~~~~~~~\Pi(g)=b(g)  a^{-1} (g),
\end{eqnarray}
where ${E_{_0}}$ is a constant matrix, and  $\Pi(g)$ defines the super Poisson structure on the $G$.
The $a(g)$  and $b(g)$ are sub-matrices of the adjoint representation of the supergroup $G$ on
$\D$ in the basis $(T_{{_a}} , {\tilde T}^{{^a}})$
\begin{eqnarray}\label{4.7}
g^{-1} T_{{_a}}~ g &=&(-1)^c ~a_{_{a}}^{^{~c}}(g) ~ T_{{_c}},\nonumber\\
g^{-1} {\tilde T}^{{^a}} g &=&
(-1)^c ~b^{^{ac}}(g)~ T_{{_c}}+{(a^{^{-st}})^{{~a}}}_{c}(g)~{\tilde T}^{{^c}}.
\end{eqnarray}
Notice that both bases $T_{{_a}} \in \G$  and ${\tilde T}^{{^a}} \in {\tG}$  satisfy relations
\eqref{2.1}-\eqref{2.3}.
We expect that there exists an equivalent T-dual $\sigma$-model in
which the roles of ${\G}$ and $\tG$ are exchanged. Accordingly, one can repeat all steps of the
previous construction to end up with the following dual $\sigma$-model
\begin{eqnarray}\label{4.8}
\tilde  S\;=\;\frac{1}{2}\int_{_{\Sigma}}\!d\sigma^{+}
d\sigma^{-}\;(-1)^{^{b}} {{\tilde  R_{_+ a}}}\;  {{\tilde E}^{^{ab}}}(\tilde g)\;{{\tilde  R_{_- b}}},
\end{eqnarray}
where ${{\tilde  R_{_\pm a}}}=(\partial_{_\pm} \tilde g ~{\tilde g}^{^{-1}})_{_a}=(-1)^{^M} \partial_{_\pm} {\tilde \phi}^{^M} {{\tilde  R_{M a}}}$
are the components of the right invariant super one-forms with ${\tilde \phi}^{^M}$ being local coordinates of  $\tilde g \in \tilde G$,
and the matrix ${{\tilde E}^{^{ab}}}(\tilde g)$ is obtained by formula
\begin{eqnarray}\label{4.9}
{{\tilde E}}(\tilde g)\;=\;\big({E_{_0}} + {\tilde \Pi}(\tilde g)\big)^{-1},
\end{eqnarray}
in which ${\tilde \Pi}(\tilde g)$ is defined as in \eqref{4.6} by replacing untilded quantities with tilded ones.
Both equations of motion of the lagrangian systems  of the models \eqref{4.4} and \eqref{4.8} can be reduced from equation of motion on the whole DSD, not depending on the choice of Manin supertriple:
\begin{eqnarray}\label{4.10}
<\partial_{\pm} l~l^{-1} , \varepsilon^{\mp}>=0,
\end{eqnarray}
where $l \in D$ \footnote{One writes $l = g {\tilde h}; g \in G, {\tilde h} \in {\tilde G}$ (such
decomposition of supergroup elements exists at least at the vicinity of the unit element); moreover, one may write
$l =  {\tilde g} h$ where  $h \in G, {\tilde g} \in {\tilde G}.$}, and subspaces $\varepsilon^{+} =span\{T_{_a} + {E_{_0}}_{_{ab}} {\tilde T}^{^{b}} \}$ and
$\varepsilon^{-} =span\{T_{_a} - {(E_{_0}^{^{st}})}_{_{ab}} {\tilde T}^{^{b}} \}$ are orthogonal with respect to
the $<. , .>$ and span the whole Lie superalgebra ${\D}$.


\section{\label{Sec.IV} Conditions that a $\sigma$-model may be given as a WZW model}

In what follows we shall obtain the conditions under which the $\sigma$-model \eqref{4.1} may be equivalent to a WZW model.
Let us first consider the standard form of WZW model based on a Lie supergroup $G$
\begin{eqnarray}\label{5.1}
S_{_{WZW}} (g)&=& \frac{1}{2} \int_{\Sigma} d\sigma^{+}
d\sigma^{-}  <g^{-1} \partial_{+}g , \; g^{-1}
\partial_{-}g >\nonumber\\
&&~~~~~~~~~~~~~~~+\frac{1}{12} \int_{B_3} <g^{-1}dg\; \hat{,}
\;[g^{-1}dg \;\hat{,}\;g^{-1}dg]>,
\end{eqnarray}
where $g^{-1}~ \partial_{\pm} g$ are the components of the left invariant super one-forms  which
may be expressed as\footnote{The relationship between the super one-forms $L^a$ and
supervector fields $V_{_b}$ is given in the usual way: $(-1)^b L^a V_{_b} ={\delta}^a_{~b}$.}
\begin{eqnarray}\label{5.2}
g^{-1}~ \partial_{_{\pm}} g =(-1)^a ~{L_{_{\pm}}^{a}}~T_{{_a}}=(-1)^a ~\partial_{_\pm} \phi^{^{M}}
\; {{_{_M}} L}^{^a}~T_{{_a}}.
\end{eqnarray}
In the second term of \eqref{5.1} the integration is over  a
$3$-dimensional (super)manifold $B_3$ which has $\Sigma$ as a boundary.
To define a WZW model, one needs a bilinear form $\Omega_{_{ab}}=<T_{{_a}} , T_{{_b}}>$ in the generators $T_{_a}$, which is non-degenerate supersymmetric
ad-invariant metric on $\G$ and satisfies in the following relation \cite{witten} (see also \cite{{ER7},{ER8}})
\begin{eqnarray}\label{5.3}
{f^c}_{ab} \;\Omega_{_{cd}}+(-1)^{bd}  {f^c}_{ad}\;\Omega_{_{cb}}\;=\;0.
\end{eqnarray}
Accordingly, the terms that are being integrated over in \eqref{5.1} can be written as
\begin{eqnarray}
<g^{-1} \partial_{+}g , \; g^{-1}
\partial_{-}g > &=& (-1)^a~\partial_{_+} \phi^{^M}  {{_{_M}} L}^{^a}~\Omega_{_{ab}}~{({L^{st})}^{^a}}_{_N}  \partial_{_-} \phi^{^N},\label{5.4}\\
<g^{-1}dg\; \hat{,}
\;[g^{-1}dg \;\hat{,}\;g^{-1}dg]> &=&  d^3 \sigma~ \varepsilon^{^{\gamma \alpha \beta}} (-1)^{^{a+bP}} \partial_{_\gamma} \phi^{^M}  {{_{_M}} L}^{^a}
\Omega_{_{ad}}\nonumber\\
&&~~~~~~~~~~~\times {f^d}_{bc} ~ {({L^{st})}^{^c}}_{_P}~ {({L^{st})}^{^b}}_{_N} ~\partial_{_\alpha} \phi^{^N} \partial_{_\beta} \phi^{^P}.\label{5.5}
\end{eqnarray}
Hence the domain of $L^a$ is implicitly generalized to $B_3$ with $\alpha= \tau, \sigma$  and $\xi$ in the second term of \eqref{5.1},
where the extra direction is labeled by $\xi$.
By regarding the action \eqref{5.1} as a $\sigma$-model action of the form
\begin{eqnarray}\label{5.6}
\bar{S}= \frac{1}{2}\int_{_{\Sigma}}\!d\sigma^+  d\sigma^- ~ \partial_{_+} \bar{\phi}^{^M}~{{_{_M}} \bar{G}}_{_{N}}~\partial_{_-} \bar{\phi}^{^N}
+ \frac{1}{12}\int_{_{B}}\!d^3\sigma~\varepsilon^{^{\gamma \beta \alpha}}
\partial_{_\gamma} \bar{\phi}^{^M}  {{_{_M}} \bar{H}}_{_{PN}} \partial_{_\alpha} \bar{\phi}^{^N} \partial_{_\beta} \bar{\phi}^{^P},
\end{eqnarray}
in which
\begin{eqnarray}\label{5.7}
\bar{H}_{_{MNP}}=(-1)^{^{M}}\;\frac{\overrightarrow{\partial}}{\partial
\bar{\phi}^{^{M}}} \bar{B}_{_{NP}}+(-1)^{^{N+M(N+P)}}\;\frac{\overrightarrow{\partial}}{\partial
\bar{\phi}^{^{N}}} \bar{B}_{_{PM}}+(-1)^{^{P(1+M+N)}}\;\frac{\overrightarrow{\partial}}{\partial
\bar{\phi}^{^{P}}} \bar{B}_{_{MN}},
\end{eqnarray}
is the torsion of the field  $\bar{B}_{_{MN}}$, and then by using equations
\eqref{5.4}  and \eqref{5.5} and finally from comparing \eqref{5.1} with \eqref{5.6}
we can read off the background of $\sigma$-model in terms of the elements of WZW model as follows:
\begin{eqnarray}
{\bar{G}}_{_{MN}} &=& (-1)^a~ {L_{_M}}^{^a}~\Omega_{_{ab}}~{({L^{st})}^{^a}}_{_N},\label{5.8}\\
\bar{H}_{_{MPN}} &=& -(-1)^{^{a+bP}} {L_{_M}}^{^a}~  \Omega_{_{ad}}  {f^d}_{bc} ~ {({L^{st})}^{^c}}_{_P}~ {({L^{st})}^{^b}}_{_N}.\label{5.9}
\end{eqnarray}
In the following we find the equation establishing a relationship between the background of $\sigma$-model
(including ${\bar{G}}_{_{MN}}$ and $\bar{B}_{_{MN}}$) and the left invariant super one-forms of Lie supergroup $G$ that the WZW model is constructed on.
For this purpose one may first use the super Maurer-Cartan equation \cite{D}
\begin{eqnarray}\label{5.10}
d  {L}^{^a}= -\frac{1}{2} (-1)^{^{bc}} {f^a}_{bc} ~ {L}^{^b} \wedge {L}^{^c},
\end{eqnarray}
to obtain
\begin{eqnarray}\label{5.11}
{({L^{st})}^{^a}}_{_M} \frac{\overleftarrow{\partial}}{\partial \phi^{^N}}
-(-1)^{^{MN}} {({L^{st})}^{^a}}_{_N} \frac{\overleftarrow{\partial}}{\partial \phi^{^M}} = (-1)^{^{N(b+M)}}
{f^a}_{bc}~ {({L^{st})}^{^c}}_{_N} {({L^{st})}^{^b}}_{_M}.
\end{eqnarray}
Then, by inserting \eqref{5.11}  into \eqref{5.9} and employing \eqref{5.8} we obtain the following {\it master equation} (ME)
\begin{eqnarray}
 {({L^{st})}^{^a}}_{_Q} ~\bar{G}^{^{QM}} {_{_M}\bar{H}}_{_{NP}} = {({L^{st})}^{^a}}_{_N} \frac{\overleftarrow{\partial}}{\partial \phi^{^P}}
-(-1)^{^{NP}} {({L^{st})}^{^a}}_{_P} \frac{\overleftarrow{\partial}}{\partial \phi^{^N}}.\label{5.12}
\end{eqnarray}
This formula will be useful in the next section.
We will solve the ME \eqref{5.12} for a given $\sigma$-model ($T$-dual models on the DSDs generated by the $(C^3+A)$)
including the metric ${\bar{G}}_{_{MN}}$ and torsion potential ${\bar{B}}_{_{MN}}$.
If there is an answer, we then obtain the left invariant super one-forms
corresponding to a Lie supergroup ${G}$ whose structure constants are still unknown.
One can then use the super Maurer-Cartan equation \eqref{5.11} together with \eqref{5.3} to determine the supergroup structure
and corresponding bilinear form which is necessary for defining a WZW model.

\section{\label{Sec.V} A hierarchy of WZW models related to super PL T-duality}

As mentioned in the introduction section, in Ref. \cite{ER8} we showed that the super PL duality relates the
$(C^3+A)$ WZW model to a $\sigma$-model defined on the $(C^3+A)$ Lie
supergroup  when the dual Lie supergroup is isomorphic to the ${C^3\otimes
{A}_{_{1,1}}}$. Then we showed that the dual model is
equivalent to a WZW model based on the $(C^3+A).i$ Lie supergroup whose Lie superalgebra is isomorphic to the $({\C}^3 +{\A})$.
Finally we announced that this process can be continued. This means that one can obtain
a hierarchy of WZW models related to the super PL T-duality in such a way that it is different
from super PL T-plurality \cite{E.nucl2020}; because as we will show below, the DSDs involved in the process are non-isomorphic.
To this end, we use the formulation presented in preceding section.
Before proceeding to obtain this hierarchy, let us give a concise review of the $(C^3+A)$ WZW model and its super PL symmetry.

\subsection{Super PL symmetry of the $(C^3 +A)$ WZW model}

As mentioned in section \ref{Sec.IV}, a key ingredient in defining the  WZW model is the bilinear form $\Omega_{_{ab}}$.
A bilinear form on the  $({\C}^3 +{\A})$ Lie superalgebra defined by (anti)commutation relations \eqref{2.4} can be obtained by \eqref{5.3},
giving
\begin{eqnarray}\label{6.1}
\Omega_{_{ab}}=\left( \begin{tabular}{cccc}
              $0$&  $1$ & 0 & $0$\\
              $1$ & 0 & 0& $0$ \\
              $0$ & 0 & 0& $1$ \\
              $0$ & $0$ & $-1$ &0\\
                \end{tabular} \right).
\end{eqnarray}
Let us introduce a supergroup element represented by
\begin{eqnarray}
g = e^{\chi T_4}~e^{y T_1}~e^{x T_2}~e^{\psi T_3}, \label{6.2}
\end{eqnarray}
where $x(\tau , \sigma)$ and $y(\tau , \sigma)$ denote bosonic fields
while $\psi(\tau , \sigma)$ and $\chi(\tau , \sigma)$ stand for fermionic fields.
To calculate the left invariant super one-forms we use \eqref{2.4} and
\eqref{5.2} together with \eqref{6.2}. Then we find that
\begin{eqnarray}\label{6.3}
{L_{_{\pm}}^{1}}&=&\partial_{\pm} y,~~~~~~~~~~~~~~~~~~~~{L_{_{\pm}}^{2}}=\partial_{\pm} x - \partial_{\pm} \chi \frac{\chi}{2},\nonumber\\
{L_{_{\pm}}^{3}}&=&-\partial_{\pm} \psi + \partial_{\pm} \chi~y,~~~~~{L_{_{\pm}}^{4}}=-\partial_{\pm} \chi.
\end{eqnarray}
Hence, by using \eqref{5.1} and \eqref{5.4} together with \eqref{5.5}, the WZW action on the $(C^3+A)$ Lie supergroup is written as \cite{ER8}
\begin{eqnarray}\label{6.4}
S_{_{WZW}} (g) =  \frac{1}{2}
\int d\sigma^{+} d\sigma^{-}\;\Big[\partial_{+} y
\partial_{-} x + \partial_{+} x
\partial_{-} y - \partial_{+} \psi\;
\partial_{-} \chi + \partial_{+} \chi\;
\partial_{-} \psi +  \partial_{+} y ~{\chi}~  \partial_{-} \chi\Big].
\end{eqnarray}
The supersymmetric part of the action gives the metric in the coordinate basis,
whereas the super antisymmetric part  gives the super antisymmetric tensor. Thus,
the corresponding line element and $B$-field in the coordinate basis are, respectively, read off
\begin{eqnarray}
ds^2 &=& 2 dy dx -2 d \psi d \chi + \chi dy d \chi,\nonumber\\
B &=& \frac{1}{2} \chi~ dy \wedge d \chi.\label{6.4.1}
\end{eqnarray}
In order to investigate the super PL symmetry of the WZW model \eqref{6.4} one has to employ equation \eqref{4.2}. First,
we need to find the left invariant supervector fields on the $(C^3 +A)$.
By applying \eqref{6.3} in relation $(-1)^b {L}^{a} {{V_b}}=\delta^a_{~b}$ we obtain
\begin{eqnarray}\label{6.5}
V_1 = \frac{ \overrightarrow{\partial}}{\partial y},~~~~~~~V_2= \frac{ \overrightarrow{\partial}}{\partial
x},~~~~~~~~V_3=\frac{ \overrightarrow{\partial}}{\partial
\psi},~   ~~~~~~V_4= \frac{\chi}{2} \frac{
\overrightarrow{\partial}}{\partial  x}+{y}\frac{
\overrightarrow{\partial}}{\partial \psi}+\frac{
\overrightarrow{\partial}}{\partial \chi}.
\end{eqnarray}
Then, by comparing actions \eqref{4.1} and \eqref{6.4}, one obtains the background matrix ${\cal E}_{_{MN}}$ corresponding to the action \eqref{6.4}.
Now, by substituting the resulting background matrix
and also the supervector fields \eqref{6.5} on the right hand side of \eqref{4.2} and then by
direct calculation of Lie derivative of ${\cal E}_{_{MN}}$ with respect to $V_a$,
one can obtain the structure constants of the dual Lie superalgebra to the $({\C}^3 +{\A})$ in
such a way that only non-zero commutation relation of the dual pair is \cite{ER8}
\begin{eqnarray}\label{6.6}
[{\tilde T}^2 , {\tilde T}^3] = -\frac{1}{2} {\tilde T}^4,
\end{eqnarray}
in which grade$({\tilde T}^1)$=grade$({\tilde T}^2)=0$ and grade$({\tilde T}^3)$=grade$({\tilde T}^4)=1$.
According to the classification of decomposable Lie superalgebras of the type $(2|2)$ \cite{ER6}, the Lie superalgebra defined by
\eqref{6.6} is isomorphic to the  ${\C}^3 \oplus {\A}_{1,1}$, which we denote by ${\C}^3 \oplus {\A}_{1,1}.ii$ \footnote{Note that
the ${\C}^3 \oplus {\A}_{1,1}.ii$ is also isomorphic to the ${\C}^3 \oplus {\A}_{1,1}.i$ of Table 1.}.
Note that both sets of generators of \eqref{2.4} and \eqref{6.6} are maximally isotropic with respect
to the bilinear form defined by the brackets \eqref{2.1}. Nevertheless,
the $(({\C}^3 +{\A})~,~ {\C}^3 \oplus {\A}_{1,1}.ii)$ as a Lie superbialgebra satisfies mixed super Jacobi identities \eqref{2.3}.

\subsection{Super PL T-dual $\sigma$-models on the superdouble $\big((C^3 +A)~,~ C^3 \otimes {A}_{1,1}.ii\big)$}
We shall construct a pair of super PL T-dual $\sigma$-models which is associated with the DSD
$(({\C}^3 +{\A})~,~ {\C}^3 \oplus {\A}_{1,1}.ii)$. Using relations \eqref{2.2}, \eqref{2.4} and \eqref{6.6} we conclude that
the Lie superalgebra of the superdouble $(({\C}^3 +{\A})~,~ {\C}^3 \oplus {\A}_{1,1}.ii)$
obeys the following set of non-trivial (anti)commutation relations\footnote{It can be easily shown that
the Lie superalgebra $(({\C}^3 +{\A})~,~ {\C}^3 \oplus {\A}_{1,1}.ii)$ with the (anti)commutation relations \eqref{6.7}
is isomorphic with each of the DSDs belonging to family DSD1 of Theorem 1.} \cite{ER8}
\begin{eqnarray}
[T_{_1} , T_{_4}] &=&T_{_3},~~~~~~~~~~~\{T_{_4} , T_{_4}\}=T_{_2},~~~~~~~~~~~~~~~~[{\tilde T}^{^2} , {\tilde T}^{^3}]=-\frac{1}{2}{\tilde T}^{^4},\nonumber\\
{[T_{_1} , {\tilde T}^{^3}]}&=& - {\tilde T}^{^4},~~~~~~~~[T_{_4} , {\tilde T}^{^2}]=-\frac{1}{2}T_{_3}-{\tilde T}^{^4},
~~~~\{T_{_4} , {\tilde T}^{^3}\}=-\frac{1}{2} T_{_2}-{\tilde T}^{^1}.\label{6.7}
\end{eqnarray}
Below, we show that the original $\sigma$-model on the $(C^3 +A)$ is the same as the WZW model obtained in \eqref{6.4}.
It can also be interesting to show that the dual model is itself equivalent to a WZW model.\\\\
{\it $\bullet$ Original model as WZW model based on the $(C^3 +A)$.}
\\
In order to construct the original $\sigma$-model with
the $(C^3+A)$ Lie supergroup as the target space, we use the same parametrization as \eqref{6.2}.
Using \eqref{4.5} we then find that
\begin{eqnarray}\label{6.8}
{R_{_{\pm}}^{1}}&=&\partial_{\pm} y,~~~~~~~~~~~~~~~~~~~~{R_{_{\pm}}^{2}}=\partial_{\pm} x + \partial_{\pm} \chi \frac{\chi}{2},\nonumber\\
{R_{_{\pm}}^{3}}&=&\partial_{\pm} y~ \chi -\partial_{\pm} \psi,~~~~~~~{R_{_{\pm}}^{4}}=-\partial_{\pm} \chi.
\end{eqnarray}
In addition, using \eqref{4.7} and the second formula of
\eqref{4.6} together with \eqref{6.2} and \eqref{6.7}, one can obtain the super Poisson structure, giving us
\begin{eqnarray}
{\Pi}^{ab}(g)=\left( \begin{tabular}{cccc}
                 $0$ & $0$ & $0$ & $0$ \\
                 $0$ & $0$  & -$\frac{\chi}{2}$& $0$ \\
                 $0$ & $\frac{\chi}{2}$& $0$ & $0$ \\
                 $0$ & $0$ & $0$ & $0$ \\
                 \end{tabular} \right).\label{6.9}
\end{eqnarray}
It is more appropriate to choose the $\sigma$-model constant matrix as follows:
\begin{eqnarray}\label{6.10}
{E_{0}}_{_{ab}}&=& \left( \begin{tabular}{cccc}
                 0 & 1 & 0 & 0  \\
                 1 & 0 & 0 & 0  \\
                  0& 0 & 0 & 1 \\
                  0& 0 & -1 & 0 \\
                 \end{tabular} \right).
\end{eqnarray}
With these relations at hand and employing \eqref{4.4} together with the first formula of \eqref{4.6}, one can get
the original $\sigma$-model which will be nothing but the WZW model \eqref{6.4}.\\\\
{\it $\bullet$  Dual model as WZW model based on the $(C^3 +A).i$.}\\
We parameterize the dual Lie supergroup $C^3 \otimes {A}_{1,1}.ii$ with
bosonic coordinates $({\tilde y}, {\tilde x})$ and fermionic ones $({\tilde \psi}, {\tilde \chi})$ so that its elements
are defined as in \eqref{6.2} by replacing untilded quantities with tilded ones.
In the same way, to construct out the dual $\sigma$-model
one needs to calculate the right invariant super one-forms ${{\tilde  R_{_\pm a}}}=(\partial_{_\pm} \tilde g ~{\tilde g}^{^{-1}})_{_a}$
on the $C^3 \otimes {A}_{1,1}.ii$. By utilizing \eqref{6.6} we obtain
\begin{eqnarray}\label{6.11}
{\tilde R}_{\pm_{1}}= {\partial_\pm}{\tilde y},~~~~ ~~~{\tilde R}_{\pm_{2}}= {\partial_\pm}{\tilde x},~~~ ~
{\tilde R}_{\pm_{3}}= {\partial_\pm}{\tilde \psi},~~~~~~~{\tilde R}_{\pm_{4}}= -{\partial_\pm}{\tilde \psi} \frac{\tilde x}{2}
+{\partial_\pm}{\tilde \chi}.
\end{eqnarray}
As mentioned in section \ref{Sec.III}, to obtain the dual super Poisson structure one must use equation \eqref{4.7}
by replacing untilded quantities with tilded ones. Thus, by employing \eqref{6.7}, we then obtain
\begin{eqnarray}
{\tilde \Pi}_{ab}(\tilde g)=\left( \begin{tabular}{cccc}
                 $0$ & $0$ & $0$ & -${\tilde \psi}$ \\
                 $0$ & $0$  & $0$& $0$ \\
                 $0$ & $0$& $0$ & $0$ \\
                 ${\tilde \psi}$ & $0$ & $0$ & -${\tilde x}$ \\
                 \end{tabular} \right).\label{6.12}
\end{eqnarray}
Finally, inserting \eqref{6.10} and \eqref{6.12} into \eqref{4.9} and then using \eqref{6.11} and \eqref{4.8},
the dual action on the $C^3 \otimes {A}_{1,1}.ii$ are worked out \cite{ER8}
\begin{eqnarray}\label{6.13}
{\tilde S} =  \frac{1}{2}
\int d\sigma^{+} d\sigma^{-}\;\Big[\partial_{+} {\tilde y}
\partial_{-} {\tilde x} + \partial_{+} {\tilde x}
\partial_{-} {\tilde y} -\partial_{+} {\tilde \psi} \partial_{-} {\tilde \chi} + \partial_{+} {\tilde \chi}
\partial_{-} {\tilde \psi} - 3 {\tilde x} \partial_{+} {\tilde \psi}\;
\partial_{-} {\tilde \psi}\Big].
\end{eqnarray}
The background corresponding to the above action including the supersymmetric metric and the super antisymmetric tensor
field  are, in the coordinate basis, read off
\begin{eqnarray}
d{\tilde s}^2 &=& 2 d {\tilde y} d {\tilde x} -2 d {\tilde \psi} d {\tilde \chi},\nonumber\\
{\tilde B} &=& -\frac{3  }{2} {\tilde x}~ d {\tilde \psi} \wedge d {\tilde \psi}.\label{6.13.1}
\end{eqnarray}
The next step in this procedure is to show that the dual $\sigma$-model constructed out on the $C^3 \otimes {A}_{1,1}.ii$
can be equivalent to a WZW model based on  $\hat{G}=(C^3 +A).i$ Lie supergroup whose Lie superalgebra is isomorphic to the $({\C}^3 +\A)$.
First, we assume that Lie supergroup $\hat{G}$ is unknown. To begin with we determine the metric and $\tilde B$-field corresponding to the action \eqref{6.13}.
By identifying the action \eqref{6.13} with the $\sigma$-model of the form \eqref{4.1} we find that only the non-zero components of the metric
and ${\tilde B}$-field are ${\tilde G}_{_{{\tilde y} {\tilde x}}}={\tilde G}_{_{{\tilde \psi} {\tilde \chi}}}=1$
and ${\tilde B}_{_{{\tilde \psi} {\tilde \psi}}}=3{\tilde x}$, respectively. One quickly finds that the only non-zero component of ${\tilde H}$
is ${\tilde H}_{_{{\tilde x}{\tilde \psi} {\tilde \psi}}}=2$.
Inserting these into the ME \eqref{5.12}, we can obtain the left invariant super one-forms of Lie supergroup $\hat{G}$. It
results\footnote{Here we have parameterized the $\hat{G}$ with the coordinates $({\hat{y}}, {\hat{x}}; {\hat{\psi}}, {\hat{\chi}})$.}
\begin{eqnarray}\label{6.14}
{\hat{L}_{_{\pm}}^{1}}&=&\partial_{\pm} \hat{y} -\frac{3}{2} \partial_{\pm} \hat{\psi}~\hat{\psi},~~~~~~~~~~~
{\hat{L}_{_{\pm}}^{2}}=\partial_{\pm} \hat{x},\nonumber\\
{\hat{L}_{_{\pm}}^{3}}&=&-\partial_{\pm} \hat{\psi},~~~~~~~~~~~~~~~~~~~~~~
{\hat{L}_{_{\pm}}^{4}}=\frac{3}{2} \partial_{\pm} \hat{x}~\hat{\psi}-\frac{3}{2} \partial_{\pm} \hat{\psi}
~ \hat{x} - \partial_{\pm} \hat{\chi}.
\end{eqnarray}
With this result at hand, one may employ the super Maurer-Cartan
equation \eqref{5.10} to obtain the structure constants of Lie superalgebra $\hat{\G}$ of the $\hat{G}$. Then, the $\hat{\G}$ is defined by
the following non-zero (anti)commutation relations
\begin{eqnarray}\label{6.15}
[\hat{T}_2 , \hat{T}_3] = -3 \hat{T}_4,~~~~~~~~~~~~~\{\hat{T}_3 , \hat{T}_3\} = 3 \hat{T}_1.
\end{eqnarray}
It can be easily shown that the $\hat{\G}$ is isomorphic to
the $({\C}^3 + {\A})$ under transformation $\hat{T}_1 \rightarrow {T}_2/3,
\hat{T}_2 \rightarrow -3{T}_1, \hat{T}_3 \rightarrow {T}_4, \hat{T}_4 \rightarrow {T}_3$, and hence we denote it by $({\C}^3 + {\A}).i$.
It is worth noting that the bilinear form on the $({\C}^3 +{\A}).i$ as a non-degenerate solution to \eqref{5.3} is obtained to be the same as
\eqref{6.1}. On the other hand, the parametrization of a general element of $({C}^3 +{A}).i$ in the form of
\begin{eqnarray}
\hat{g} = e^{(\hat{\chi} +\frac{3}{2} \hat{x}
\hat{\psi}) \hat{T}_4}~e^{\hat{y} \hat{T}_1}~e^{\hat{x} \hat{T}_2}~e^{\hat{\psi} \hat{T}_3}, \label{6.16}
\end{eqnarray}
leads to the super one-forms \eqref{6.14}.
Finally, employing \eqref{5.1} and using the above results, one can get the WZW action on the $({C}^3 + {A}).i$, giving us
\begin{eqnarray}\label{6.17}
{\hat{S}}_{_{WZW}} (\hat{g}) =  \frac{1}{2}
\int d\sigma^{+} d\sigma^{-}\Big[\partial_{+} {\hat y}
\partial_{-} {\hat x} + \partial_{+} {\hat x}
\partial_{-} {\hat y} -\partial_{+} {\hat \psi} \partial_{-} {\hat \chi} + \partial_{+} {\hat \chi}
\partial_{-} {\hat \psi} - 3 {\hat x} \partial_{+} {\hat \psi}\;
\partial_{-} {\hat \psi}\Big].
\end{eqnarray}
which is nothing but \eqref{6.13}. Thus, our results show that the dual $\sigma$-model to the $(C^3 +A)$ WZW model is itself a WZW model.
In fact, we showed that the $(C^3 +A)$ WZW model does remain invariant under the super PL T-duality transformation, that is, the model is super PL self-dual.

\subsection{Super PL symmetry of the $(C^3 +A).i$ WZW model}
Here we shall investigate super PL symmetry in the $(C^3 +A).i$ WZW model describing by action \eqref{6.17}.
From the comparison of action \eqref{6.17} with \eqref{4.1}, one can write the background matrix of the model as follows:
\begin{eqnarray}
{\hat{\cal E}}_{_{MN}}=\left( \begin{tabular}{cccc}
                 $0$ & $1$ & $0$ & $0$ \\
                 $1$ & $0$  & $0$& $0$ \\
                 $0$ & $0$& $3 { \hat{x}}$ & $1$ \\
                 $0$ & $0$ & -$1$ & $0$ \\
                 \end{tabular} \right).\label{6.18}
\end{eqnarray}
Furthermore, we need to calculate the left-invariant supervector fields on the $(C^3 +A).i$. Using \eqref{6.14} one gets
\begin{eqnarray}\label{6.19}
\hat{V}_1 = \frac{ \overrightarrow{\partial}}{\partial \hat{y}},~~~~~~~\hat{V}_2= \frac{ \overrightarrow{\partial}}{\partial
\hat{x}} +\frac{3}{2} \hat{\psi} \frac{
\overrightarrow{\partial}}{\partial \hat{\chi}},~~~~~~~~
\hat{V}_3=\frac{3}{2} \hat{\psi} \frac{
\overrightarrow{\partial}}{\partial \hat{y}} + \frac{
\overrightarrow{\partial}}{\partial \hat{\psi}}-\frac{3}{2} \hat{x} \frac{
\overrightarrow{\partial}}{\partial \hat{\chi}},~~~~~\hat{V}_4= \frac{
\overrightarrow{\partial}}{\partial \hat{\chi}}.
\end{eqnarray}
Substituting the above relations in formula \eqref{4.2}
one can obtain the structure constants of the dual pair to the $(\C^3 +\A).i$ Lie
superalgebra, giving
\begin{eqnarray}\label{6.20}
\{{\tilde {\hat{T}}}^4 , {\tilde {\hat{T}}}^4\} = 3 {\tilde {\hat{T}}}^2,
\end{eqnarray}
The Lie superalgebra defined by
\eqref{6.20} is isomorphic to the  $(2{\A}_{1,1} + 2 {\A}\big)^0.i$ Lie superalgebra of Table 1.
Based on this, we denote it by $(2{\A}_{1,1} + 2 {\A}\big)^0.ii$.
Both sets of generators of \eqref{2.4} and \eqref{6.20} are maximally isotropic with respect
to the bilinear form defined by the brackets \eqref{2.1}. Nevertheless,
the $(({\C}^3 +{\A}).i , (2{\A}_{1,1} + 2 {\A}\big)^0.ii)$ as a Lie superbialgebra satisfies mixed super Jacobi identities \eqref{2.3}.

\subsection{Super PL T-dual $\sigma$-models on the superdouble $\big((C^3 +A).i , (2{A}_{1,1} + 2 {A}\big)^0.ii\big)$}

The Lie superalgebra of the superdouble $(({\C}^3 +{\A}).i , (2{\A}_{1,1} + 2 {\A}\big)^0.ii)$
obeys the following set of non-trivial (anti)commutation relations\footnote{Notice that
the Lie superalgebra $(({\C}^3 +{\A}).i , (2{\A}_{1,1} + 2 {\A}\big)^0.ii)$ with the (anti)commutation relations \eqref{6.21}
is isomorphic with each of the DSDs belonging to family ${DSD 23}^{^k}$ of Theorem 1.} \cite{ER8}
\begin{eqnarray}
[\hat{T}_{_2} , \hat{T}_{_3}] &=& -3\hat{T}_{_4},~~~~~~~~~~~~~~~~~\{\hat{T}_{_3} , \hat{T}_{_3}\}=3\hat{T}_{_1},
~~~~~~~~~~~~[{\tilde {\hat{T}}}^{^4} , {\tilde {\hat{T}}}^{^4}]=3{\tilde {\hat{T}}}^{^2},\nonumber\\
{[\hat{T}_{_2} , {\tilde {\hat{T}}}^{^4}]}&=& 3(- \hat{T}_{_4} +{\tilde {\hat{T}}}^{^3}),~~~~~~~~
[\hat{T}_{_3} , {\tilde {\hat{T}}}^{^1}]=-3 {\tilde {\hat{T}}}^{^3},
~~~~~~~~~\{\hat{T}_{_3} , {\tilde {\hat{T}}}^{^4}\}=3 {\tilde {\hat{T}}}^{^2}.\label{6.21}
\end{eqnarray}
Below, we construct $T$-dual $\sigma$-models on the DSD defined by \eqref{6.21}.
As expected, the original $\sigma$-model will be equivalent to the WZW model based on the $(C^3 +A).i$, i.e., action \eqref{6.17}.
The interesting thing about the dual model is that it will also be equivalent to a WZW model.
\\\\
{\it $\bullet$ Original model as WZW model based on the $(C^3 +A).i$.}
\\
In order to construct the original $\sigma$-model with
the $(C^3+A).i$ Lie supergroup as the target superspace, we use the same parametrization as \eqref{6.16}.
Using \eqref{4.5} we then find that
\begin{eqnarray}\label{6.22}
{\hat{R}_{_{\pm}}^{1}}&=&\partial_{\pm} \hat{y} +\frac{3}{2} \partial_{\pm} \hat{\psi} ~\hat{\psi},~~~~~~~~~~~~~~
{\hat{R}_{_{\pm}}^{2}}=\partial_{\pm} \hat{x},\nonumber\\
{\hat{R}_{_{\pm}}^{3}}&=&-\partial_{\pm} \hat{\psi},~~~~~~~~~~~~~~~~~~~~~~~~~~
{\hat{R}_{_{\pm}}^{4}}=-\frac{3}{2} \partial_{\pm} \hat{x} ~ \hat{\psi} +\frac{3}{2}\partial_{\pm} \hat{\psi}~ \hat{x} -\partial_{\pm} \hat{\chi}.
\end{eqnarray}
It follows from \eqref{4.7} and the second formula of
\eqref{4.6} that only non-zero component of the super Poisson structure is ${\hat{\Pi}}^{^{\hat{\chi} \hat{\chi}}}(\hat{g}) = -3 \hat{x}$.
By considering the $\sigma$-model constant matrix in the form of \eqref{6.10}
and by employing \eqref{4.4} together with the first formula of \eqref{4.6}, one can get
the action of $\sigma$-model on the $(C^3+A).i$ which will be nothing but the WZW model \eqref{6.17}.\\\\
{\it $\bullet$  Dual model as WZW model based on the $(C^3 +A).ii$.}\\
In order to calculate the components of the right invariant super
one-forms ${{\tilde  {\hat{R}}_{_\pm a}}}$ on the $(2{A}_{1,1} + 2 {A}\big)^0.ii$ Lie supergroup we parameterize an element of the supergroup as
\begin{eqnarray}\label{6.23}
{\tilde {\hat g}}~=~e^{\tilde {\hat{\chi}} {\tilde {\hat{T}}}^{^4}} e^{ \tilde {\hat{y}} {\tilde {\hat{T}}}^{^1}}~e^{\tilde {\hat{x}} {\tilde {\hat{T}}}^{^2}}
~e^{ \tilde {\hat{\psi}} {\tilde {\hat{T}}}^{^3}}.
\end{eqnarray}
Now, one may use equation \eqref{4.5} to get ${\tilde {\hat{R}}}_{\pm_{a}}$'s on the $(2{A}_{1,1} + 2 {A}\big)^0.ii$, giving
\begin{eqnarray}\label{6.24}
{\tilde {\hat{R}}}_{\pm_{1}}= {\partial_\pm}{\tilde {\hat{y}}},~~~~ ~~~{\tilde {\hat{R}}}_{\pm_{2}}= {\partial_\pm}{\tilde {\hat{x}}}+\frac{3}{2}
{\partial_\pm}{\tilde {\hat{\chi}}} ~{\tilde {\hat{\chi}}},~~~~
{\tilde {\hat{R}}}_{\pm_{3}}= {\partial_\pm}{\tilde {\hat{\psi}}},~~~~~~~{\tilde {\hat{R}}}_{\pm_{4}}= {\partial_\pm}{\tilde {\hat{\chi}}}.
\end{eqnarray}
To obtain the dual super Poisson structure one must employ equations \eqref{6.21} and \eqref{6.23} in formula \eqref{4.7} and the second formula of \eqref{4.6}.
It then results
\begin{eqnarray}
{\tilde {\hat \Pi}}_{ab}(\tilde {\hat g})=\left( \begin{tabular}{cccc}
                 $0$ & $0$ & $0$ & $0$ \\
                 $0$ & $0$  & $3 {\tilde {\hat{\chi}}}$& $0$ \\
                 $0$ & $-3{\tilde {\hat{\chi}}}$& $-3 {\tilde {\hat{y}}}$ & $0$ \\
                 $0$ & $0$ & $0$ & $0$ \\
                 \end{tabular} \right).\label{6.25}
\end{eqnarray}
Thus, one may insert \eqref{6.10} and \eqref{6.25} into \eqref{4.9} and then use \eqref{6.24} and \eqref{4.8} to find
the dual action on the $(2{A}_{1,1} + 2 {A}\big)^0.ii$. Using
integrating by parts, the action becomes
\begin{eqnarray}\label{6.26}
{\tilde {\hat S}} =  \frac{1}{2}
\int d\sigma^{+} d\sigma^{-}\;\Big[\partial_{+} {\tilde {\hat{y}}}
\partial_{-} {\tilde {\hat{x}}} + \partial_{+} {\tilde {\hat{x}}}
\partial_{-} {\tilde {\hat{y}}} +3 \partial_{+} {\tilde {\hat{y}}}   {\tilde {\hat{\chi}}} \partial_{-} {\tilde {\hat{\chi}}}+6
 \partial_{+} {\tilde {\hat{\chi}}} ~{\tilde {\hat{\chi}}} \partial_{-} {\tilde {\hat{y}}}
-\partial_{+} {\tilde {\hat{\psi}}} \partial_{-} {\tilde {\hat{\chi}}} + \partial_{+} {\tilde {\hat{\chi}}}
\partial_{-} {\tilde {\hat{\psi}}}\Big].~~
\end{eqnarray}
Now, one may use formulae \eqref{4.1.2} and \eqref{4.1.3} to obtain the supersymmetric metric and the super antisymmetric tensor
field corresponding to the above action in the coordinate basis, giving us
\begin{eqnarray}
d{\tilde {\hat s}}^2 &=& 2 d {\tilde {\hat y}} d {\tilde {\hat x}} -2 d {\tilde  {\hat \psi} } d {\tilde {\hat  \chi} } 
- 3  {\tilde {\hat  \chi} }  d {\tilde {\hat y}} d {\tilde {\hat  \chi} },\nonumber\\
{\tilde {\hat B}} &=& \frac{9 }{2} {\tilde {\hat  \chi} }~ d {\tilde {\hat y}} \wedge  d {\tilde {\hat  \chi} }.\label{6.26.1}
\end{eqnarray}
In order to complete the chain of WZW models related to super PL T-duality,
we show that the dual $\sigma$-model on the  $(2{A}_{1,1} + 2 {A}\big)^0.ii$
can be also equivalent to a WZW model which is built on  $\bar{G}=(C^3 +A).ii$ Lie supergroup whose Lie superalgebra is isomorphic to the $({\C}^3 +\A)$.
Similarly to the previous case, assume that Lie supergroup $\bar{G}$ is unknown.
To determine the structure of $\bar{G}$ Lie supergroup, first we find the super one-forms corresponding to $\bar{G}$.
To this end, one must insert the inverse of metric and the field strength corresponding to the dual action \eqref{6.26} into the the ME \eqref{5.12}.
From the comparison of action \eqref{6.26} with $\sigma$-model \eqref{4.1}, we find that the inverse of metric is
\begin{eqnarray}
{\tilde {\hat G}}^{{MN}}=\left( \begin{tabular}{cccc}
                 $0$ & $1$ & $0$ & $0$ \\
                 $1$ & $0$  & -$\frac{3}{2} {\tilde {\hat{\chi}}}$& $0$ \\
                 $0$ & -$\frac{3}{2} {\tilde {\hat{\chi}}}$& $0$ & $1$ \\
                 $0$ & $0$ & -$1$ & $0$ \\
                 \end{tabular} \right),\label{6.27}
\end{eqnarray}
and as the only non-zero component of ${\tilde {\hat B}}$-field is ${\tilde {\hat B}}_{_{{\tilde {\hat{y}}} {\tilde {\hat{\chi}}}}} =9{\tilde {\hat{\chi}}}/2$, the only
non-zero component of field strength $\tilde {\hat H}$ is ${\tilde {\hat H}}_{_{ {\tilde {\hat{y}}}{\tilde {\hat{\chi}}}{\tilde {\hat{\chi}}} }}=9$.
 Inserting these into the ME \eqref{5.12}, the left invariant super one-forms on the $\bar{G}$ are worked out to be of the form
\begin{eqnarray}\label{6.28}
{\bar{L}_{_{\pm}}^{1}}&=&\partial_{\pm} \bar{y},~~~~~~~~~~~~~~~~~~~~~~~~~~~~~~~~~~~~~
{\bar{L}_{_{\pm}}^{2}}=\partial_{\pm} \bar{x} -\frac{9}{2} \partial_{\pm} \bar{\chi}~\bar{\chi},\nonumber\\
{\bar{L}_{_{\pm}}^{3}}&=&-6 \partial_{\pm} \bar{y}~\bar{\chi}- \partial_{\pm} \bar{\psi}+3 \partial_{\pm} \bar{\chi}~\bar{y},~~~~~~~~
{\bar{L}_{_{\pm}}^{4}}=-\partial_{\pm} \bar{\chi}.
\end{eqnarray}
Here we have parameterized the $\bar{G}$ by the coordinates $({\bar{y}}, {\bar{x}}; {\bar{\psi}}, {\bar{\chi}})$.
By substituting the super one-forms \eqref{6.28} into
equation of the super Maurer-Cartan \eqref{5.10} one can obtain the structure constants of Lie superalgebra $\bar{\G}$ whose
non-zero superbrackets read
\begin{eqnarray}\label{6.29}
[\bar{T}_1 , \bar{T}_4] = 9 \bar{T}_3,~~~~~~~~~~~~~\{\bar{T}_4 , \bar{T}_4\} = 9 \bar{T}_2.
\end{eqnarray}
Obviously, the $\bar{\G}$ is isomorphic to
the $({\C}^3 + {\A})$ and hence we denote it by $({\C}^3 + {\A}).ii$. By using \eqref{5.3} and \eqref{6.29},
it can be shown that the bilinear form on the $({\C}^3 +{\A}).ii$ is nothing but \eqref{6.1}.
Notice that now that the structure of the Lie superalgebra $({\C}^3 +{\A}).ii$ is known,
one can obtain the super one-forms \eqref{6.28} by a convenient parametrization of the supergroup element in the following form
\begin{eqnarray}
\bar{g} = e^{\bar{\chi} \bar{T}_4}~e^{\bar{y} \bar{T}_1}~e^{\bar{x} \bar{T}_2}~e^{(\bar{\psi} +6 \bar{y} \bar{\chi}) \bar{T}_3}. \label{6.30}
\end{eqnarray}
Finally, using these results one can construct the WZW model based on the $({C}^3 + {A}).ii$
whose action is
\begin{eqnarray}
{\bar{S}}_{_{WZW}} (\bar{g})&=&  \frac{1}{2}
\int d\sigma^{+} d\sigma^{-}\;\Big[\partial_{+} {{\bar{y}}}
\partial_{-} {\ {\bar{x}}} + \partial_{+} {{\bar{x}}}
\partial_{-} { {\bar{y}}} +3 \partial_{+} { {\bar{y}}}   { {\bar{\chi}}} \partial_{-} { {\bar{\chi}}}+6
 \partial_{+}  {\bar{\chi}} ~{ {\bar{\chi}}} \partial_{-} { {\bar{y}}}\nonumber\\
&&~~~~~~~~~~~~~~~~~~~~~~~~~~~~~~~~~~~~~~~~~~~~~~~~~~~
-\partial_{+} { {\bar{\psi}}} \partial_{-} { {\bar{\chi}}} + \partial_{+} { {\bar{\chi}}}
\partial_{-} { {\bar{\psi}}}\Big].\label{6.31}
\end{eqnarray}
As it can be seen, the above action is exactly equal to the action of $\sigma$-model \eqref{6.26}.
In fact, we showed that the dual $\sigma$-model to the $(C^3 +A).i$ WZW model is itself a WZW model.

\subsection{Super PL symmetry of the $(C^3 +A).ii$ WZW model}

In order to investigate the super PL symmetry of the $(C^3 +A).ii$ WZW model, we employ formula \eqref{4.2} to find the structure constants of
the dual Lie supergroup to the $(C^3 +A).ii$. In this way, we need to specify the background matrix
of the $(C^3 +A).ii$ WZW model, as well as the left-invariant supervector fields corresponding to the $(C^3 +A).ii$.
By identifying the action \eqref{6.31} with the $\sigma$-model
of the form \eqref{4.1} we can read off the background matrix
${\bar{\cal E}}_{_{MN}}$ in the coordinate base $(d \bar{y}, d \bar{x}; d \bar{\psi}, d \bar{\chi})$ as
\begin{eqnarray}
{\bar{\cal E}}_{_{MN}}=\left( \begin{tabular}{cccc}
                 $0$ & $1$ & $0$ & $3 { {\bar{\chi}}}$ \\
                 $1$ & $0$  & $0$& $0$ \\
                 $0$ & $0$& $0$ & $1$ \\
                 -$6 { {\bar{\chi}}}$ & $0$ & -$1$ & $0$ \\
                 \end{tabular} \right).\label{6.32}
\end{eqnarray}
Furthermore, by applying \eqref{6.28} in relation $(-1)^b {L}^{a} {{V_b}}=\delta^a_{~b}$ one can obtain
the left-invariant supervector fields on the $(C^3 +A).ii$, giving us
\begin{eqnarray}\label{6.33}
\bar{V}_1 = \frac{ \overrightarrow{\partial}}{\partial \bar{y}} -6 { {\bar{\chi}}} \frac{ \overrightarrow{\partial}}{\partial \bar{\psi}},~~~~~~~
\bar{V}_2= \frac{ \overrightarrow{\partial}}{\partial \bar{x}},~~~~~~~~
\bar{V}_3= \frac{\overrightarrow{\partial}}{\partial \bar{\psi}} ,~~~~~
\bar{V}_4= \frac{9}{2} { {\bar{\chi}}}  \frac{ \overrightarrow{\partial}}{\partial \bar{x}} +
3 { {\bar{y}}}  \frac{ \overrightarrow{\partial}}{\partial \bar{\psi}} +\frac{ \overrightarrow{\partial}}{\partial \bar{\chi}}.
\end{eqnarray}
Now, by substituting relations \eqref{6.32} and \eqref{6.33} in formula \eqref{4.2}
one can obtain the structure constants of the dual pair to the $(\C^3 +\A).ii$ Lie
superalgebra in such a way that only non-zero commutation relation of the dual pair is
\begin{eqnarray}\label{6.34}
[{\tilde {\bar{T}}}^2 , {\tilde {\bar{T}}}^3]=- \frac{9}{2} {\tilde {\bar{T}}}^4.
\end{eqnarray}
Actually, the resulting Lie superalgebra is isomorphic to the ${\C}^3 \oplus {\A}_{1,1}.i$ of Table 1, and  hence we denote this dual pair
by ${{\C}^3 \oplus {\A}_{1,1}}.iii$. Indeed, two sets of generators \eqref{6.29} and
\eqref{6.34} are dual to each other in the sense of \eqref{2.1}. Despite this, one can check that the $(\C^3 +\A).ii$ Lie
superalgebra and its dual pair, the ${{\C}^3 \oplus {\A}_{1,1}}.iii$, satisfy mixed super Jacobi identities \eqref{2.3}.
The Lie superalgebra of the DSD which we refer to as the $\left((\C^3 +\A).ii , {{\C}^3 \oplus {\A}_{1,1}}.iii\right)$
is defined by the following non-zero (anti)commutation relations
\begin{eqnarray}
{[\bar{T}_1 , \bar{T}_4]} &=& 9 \bar{T}_3,~~~~~~\{\bar{T}_4 , \bar{T}_4\} = 9 \bar{T}_2,~~~~[{\tilde {\bar{T}}}^2 , {\tilde {\bar{T}}}^3]=- \frac{9}{2} {\tilde {\bar{T}}}^4,~~~~~
[\bar{T}_1 , {\tilde {\bar{T}}}^3]=-9 {\tilde {\bar{T}}}^4,\nonumber\\
{[\bar{T}_{_4} , {\tilde {\bar T}}^{^2}]}&=&-\frac{9}{2}\bar{T}_{_3}-9 {\tilde {\bar T}}^{^4},
~~~~\{\bar{T}_{_4} , {\tilde {\bar T}}^{^3}\}=-\frac{9}{2} \bar{T}_{_2}-9{\tilde {\bar T}}^{^1}.\label{6.35}
\end{eqnarray}
It is interesting to comment on the fact that the DSDs $\left((\C^3 +\A).ii , {{\C}^3 \oplus {\A}_{1,1}}.iii\right)$ and
$\big((\C^3 +\A) $,$ {{\C}^3 \oplus {\A}_{1,1}}.ii\big)$
as Lie superalgebras are isomorphic, and
so one can find an isomorphism that preserves also the bilinear
form $<.\;,\;.>$, so that they belong  to the same DSD.
The isomorphism of Manin supertriples between $\left((\C^3 +\A).ii , {{\C}^3 \oplus {\A}_{1,1}}.iii\right)$  and
$\left((\C^3 +\A) , {{\C}^3 \oplus {\A}_{1,1}}.ii\right)$ is given by the following
transformations
\begin{eqnarray}\label{6.36}
&&\bar{T}_1 \rightarrow  9 {T}_1,~~~~~~\bar{T}_2 \rightarrow \frac{1}{9} T_2,~~~~~~\bar{T}_3 \rightarrow  T_3,~~~~~~\bar{T}_4 \rightarrow  T_4,\nonumber\\
&&\tilde {\bar{T}}^1 \rightarrow  \frac{1}{9} {\tilde T}^1,~~~~~~ \tilde {\bar{T}}^2 \rightarrow  {9} {\tilde T}^2,~~~~~~\tilde {\bar{T}}^3 \rightarrow  {\tilde T}^3,~~~~~~
\tilde {\bar{T}}^4 \rightarrow  {\tilde T}^4.
\end{eqnarray}
According to the above result, the story of hierarchy of WZW models related to super PL T-duality will be again repeated.

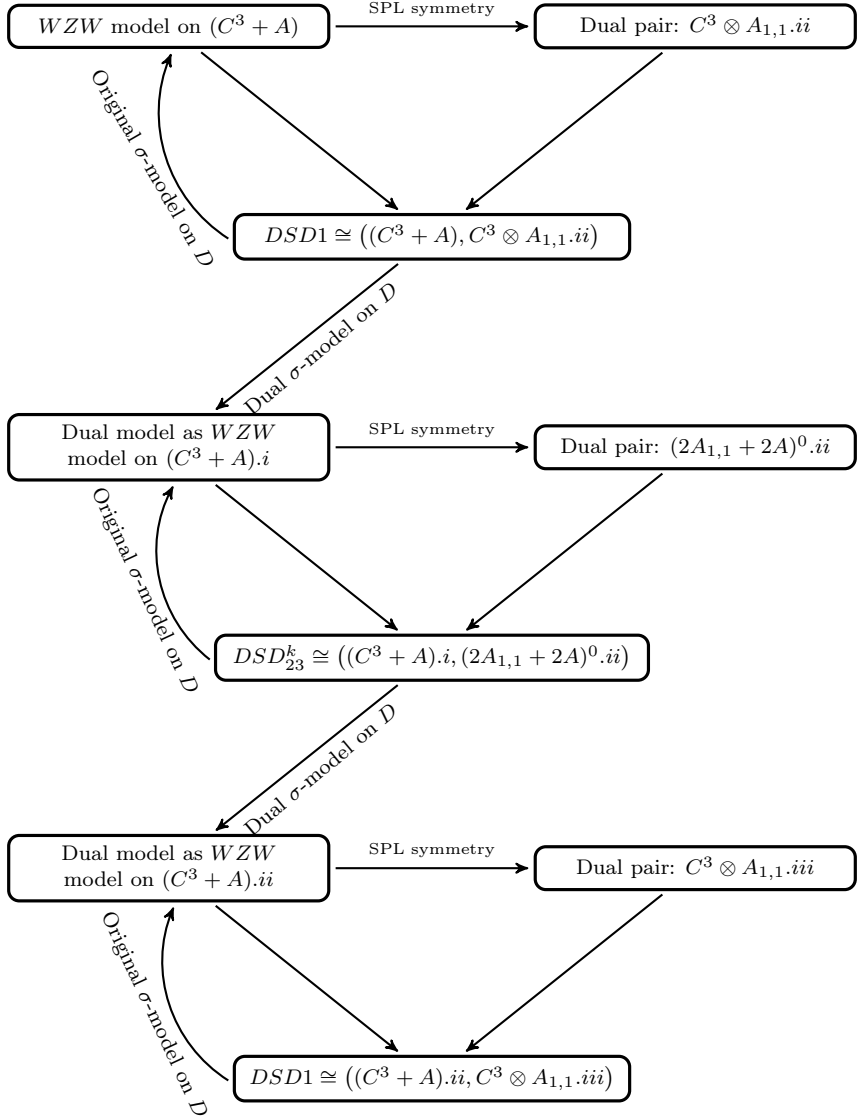
\begin{figure}[H]
\centering
\begin{tikzpicture}[scale=0.70, font=\tiny]
	\path (-5,0) node[format]  (1) {\parbox{4cm}
		{
			\centering $WZW$ model on $(C^3+A)$
		}
	};
		\path (5,0) node[format]  (2) {\parbox{4cm}
		{
			\centering Dual pair: $C^3 \otimes A_{1,1}.ii$
		}
	};
	
 \path[arrow,->] (1) edge            node[midway,above ]  {SPL symmetry}     (2);

\path (0,-4) node[format]  (3) {\parbox{5cm}
 	{
 		\centering $DSD1 \cong \big((C^3+A),C^3 \otimes A_{1,1}.ii\big)$
 	}
 };
  \path[arrow,->] (1) edge            node[midway,above ]  {}     (3);
  \path[arrow,->] (2) edge            node[midway,above ]  {}     (3);
  \path[arrow,->,bend left=40] (3) edge            node[midway,rotate=-60,above ]  { {  }}node[midway,rotate=-60,below ]  { { \scriptsize {Original $\sigma$-model on $D$}}}     (1);

    \path (-5,-8) node[format]  (4) {\parbox{4cm}
    	{
    		\centering Dual model as $WZW$\\
    		model on $(C^3+A).i$
    	}
    };
    \path (5,-8) node[format]  (5) {\parbox{4cm}
    	{
    		\centering Dual pair: $(2A_{1,1}+2A)^0.ii$
    	}
    };
    	\path[arrow,->] (3) edge            node[midway,rotate=40,above ]  {}
    	node[midway,rotate=40,below ]  {{ \scriptsize Dual $\sigma$-model on $D$}}     (4);
	\path[arrow,->] (4) edge            node[midway,above ]  {SPL symmetry}     (5);
	
	\path (0,-12) node[format]  (6) {\parbox{5.5cm}
		{
			\centering $DSD^k_{23} \cong \big((C^3+A).i,(2A_{1,1} + 2A)^0.ii\big)$
		}
	};
	\path[arrow,->] (4) edge            node[midway,above ]  {}     (6);
	\path[arrow,->] (5) edge            node[midway,above ]  {}     (6);
	\path[arrow,->,bend left=40] (6) edge            node[midway,rotate=-70,above ]  { {  }}node[midway,rotate=-65,below ]  { { \scriptsize {Original $\sigma$-model on $D$}}}     (4);

	    \path (-5,-16) node[format]  (7) {\parbox{4cm}
		{
			\centering Dual model as $WZW$\\
			model on $(C^3+A).ii$
		}
	};
	\path (5,-16) node[format]  (8) {\parbox{4cm}
		{
			\centering Dual pair: $C^3\otimes A_{1,1}.iii$
		}
	};
	\path[arrow,->] (6) edge            node[midway,rotate=40,above ]  {}
								node[midway,rotate=40,below ]  {{ \scriptsize Dual $\sigma$-model on $D$}}   (7);
	\path[arrow,->] (7) edge            node[midway,above ]  {SPL symmetry}     (8);
	
	\path (0,-20) node[format]  (9) {\parbox{5cm}
		{
			\centering $DSD1 \cong \big((C^3+A).ii,C^3\otimes A_{1,1}.iii\big)$
		}
	};
	
	\path[arrow,->] (7) edge            node[midway,above ]  {}     (9);
	\path[arrow,->] (8) edge            node[midway,above ]  {}     (9);
	\path[arrow,->,bend left=40] (9) edge            node[midway,rotate=-65,above ]  { {  }}node[midway,rotate=-65,below ]  { { \scriptsize {Original $\sigma$-model on $D$}}}     (7);
\end{tikzpicture}
\caption{Scheme of a hierarchy of WZW models related to super PL T-duality.} \label{fig}
\end{figure}

In summary,  starting the $(C^3+A)$ WZW model we showed that the dual $\sigma$-model to the $(C^3 +A)$ WZW model was itself a WZW model,
which was built on the $(C^3+A).i$. Then, investigating the super PL symmetry in the $(C^3+A).i$ WZW model,
we showed that the dual $\sigma$-model to the $(C^3 +A).i$ WZW model was again a WZW model, which was built this time on the $(C^3+A).ii$.
Thus, we found a hierarchy of WZW models related to super PL T-duality so that
it was different from super PL T-plurality; because as we showed, the DSDs involved in the process were non-isomorphic.
For the sake of clarity, the results obtained in this section
are depicted as a flowchart in Figure \ref{fig}.
\section{\label{Sec.VI}  Summary and concluding remarks}

We have classified $(4|4)$-dimensional DSDs generated by the Manin supertriples
with the first sub-superalgebra $({\C}^3 + {\A})$, i.e.,  Lie superalgebras provided with
bilinear non-degenerate supersymmetric and super ad-invariant form such
that they can be decomposed into a pair of maximally isotropic sub-superalgebras.
We claimed that there are just 24 classes of the non-isomorphic DSDs.
The result has been summarized in Theorem 1.
In section \ref{Sec.IV} we introduced a general formulation to find the conditions under which a two-dimensional $\sigma$-model
may be equivalent to a WZW model.
With the help of this formulation, we have found a chain of WZW models related to the super PL T-duality:
we showed that the super PL T-duality relates the
$(C^3+A)$ WZW model to a $\sigma$-model defined on the $(C^3+A)$ Lie
supergroup  when the dual Lie supergroup is ${C^3\otimes
{A}_{_{1,1}}}.ii$. Then we showed that the dual model is
equivalent to a WZW model based on the $(C^3+A).i$ Lie supergroup whose Lie superalgebra is isomorphic to the $({\C}^3 +{\A})$ \cite{ER8}.
Accordingly, we concluded that the $(C^3+A)$ WZW model
did remain invariant under the
super PL T-duality transformation, that is, the model was super PL self-dual.
As the same way, we have shown that the super PL T-duality relates the
$(C^3+A).i$ WZW model to a $\sigma$-model defined on the dual Lie supergroup $(2A_{1,1} + 2A)^0.ii$.
We have also stressed that the dual model built on the $(2A_{1,1} + 2A)^0.ii$ is equivalent to the WZW model on isomorphic Lie supergroup $(C^3+A).ii$.
Finally, we described a pair of super PL T-dual $\sigma$-models which was associated with the
$DSD1 \cong \big((C^3+A).ii,C^3\otimes A_{1,1}.iii\big)$.
Note that in this process, the DSDs $DSD1 \cong \big((C^3+A) , C^3\otimes A_{1,1}.ii\big)$
and $DSD^k_{23} \cong \big((C^3+A).i , (2A_{1,1} + 2A)^0.ii\big)$ were non-isomorphic, as mentioned in Theorem 1.
Accordingly, this process was different from super PL T-plurality.
Because of being isomorphism the DSDs $\big((C^3+A) , C^3\otimes A_{1,1}.ii\big)$ and $\big((C^3+A).ii , C^3\otimes A_{1,1}.iii\big)$,
the process of calculating chain of WZW models was repeated again.

As mentioned in the introduction section, we have already obtained the dual $\sigma$-model to the $GL(1|1)$ WZW model \cite{ER7},
as well as to the Heisenberg WZW model \cite{eghbali11} and the $GL(2 , \mathds{R})$ WZW model \cite{Exact}. In all cases, the dual $\sigma$-models were conformally invariant up to one-loop order and in some cases even up to two-loops,
but they were not equivalent to a WZW model.
Therefore, finding an example like the $(C^3+A)$ WZW model is extremely rare.

\subsection*{Acknowledgements}
This work was supported by Iran National Science Foundation: INSF  under research fund No. 94/s/41258.

\appendix

\section{Isomorphism transformation between the Manin supertriples with the first sub-superalgebra ${\cal G} \equiv ({\C}^3 + {\A})$}

In this appendix we give isomorphism transformation between the Manin supertriples
with the first sub-superalgebra ${\cal G} \equiv ({\C}^3 + {\A})$ of Table 1 that
are isomorphic as Lie superalgebras. The DSDs $\D$ and $\D'$ with these special bases
$X_{_A} = (T_{_1}, T_{_2}, {\tilde T}^{^1}, {\tilde T}^{^2}; T_{_3}, T_{_4}, {\tilde T}^{^3}, {\tilde T}^{^4}),
X'_{_A} = (T'_{_1}, T'_{_2}, {\tilde T}'^{^1}, {\tilde T}'^{^2}; T'_{_3}, T'_{_4},\\ {\tilde T}'^{^3}, {\tilde T}'^{^4})$
are isomorphic iff there is a superinvertible $8 \times 8$ transformation matrix ${\mathbb{C}_{_{A}}}^{^B}$
such that the linear map given by
equation \eqref{2.5} transforms the Lie multiplication of $\D$ into that of $\D'$  and preserves
the bilinear form \eqref{2.7} as well as equation \eqref{2.6}.
As we will show, the isomorphism transformations between the Manin supertriples of Table 1
are not unique; they contain several free parameters.
One can give them in a simple form setting the parameters zero or
one.

\subsection{Structure of $DSD 1:~\big({\cal G} , {\cal {I}}_{_{(2|2)}}\big) \cong
\big({\cal G} , {\C}^3 \oplus {{\A}_{_{1,1}}}.i\big) \cong  \big({\cal G} , {({\C}^3 + {\A})_{_{k=4}}}^{\hspace{-5mm}\epsilon=-1}\big)$}

$\bullet$~The isomorphism of Manin supertriples between $\big({\cal G} , {\cal {I}}_{_{(2|2)}}\big)$ and
$\big({\cal G} , {\C}^3 \oplus {{\A}_{_{1,1}}}.i\big)$ is given by the following transformation
\begin{eqnarray}\label{A.1}
X'_{_1} & = & \frac{1}{ab} X_{_1},~~~~~~~~~~~~~~~~~~~~ X'_{_5} = -\frac{1}{b} X_{_5}, \nonumber\\
X'_{_2} & = & a^{^2}  X_{_2},~~~~~~~~~~~~~~~~~~~~~X'_{_6} = -a  X_{_6},\nonumber\\
X'_{_3} & = & a^{^2}  X_{_2}+ ab X_{_3},~~~~~~~~~~~X'_{_7} = -c X_{_5} -b  X_{_7},\nonumber\\
X'_{_4} & = & -\frac{1}{ab} X_{_1}+\frac{1}{a^{^2}} X_{_4},~~~~~~~X'_{_8} = -\frac{1}{a} X_{_8},
\end{eqnarray}
where $(X_{_1},\cdots, X_{_8})$ and $(X'_{_1},\cdots, X'_{_8})$ are generators of the Manin supertriples $\big({\cal G} , {\cal {I}}_{_{(2|2)}}\big)$ and
$\big({\cal G} , {\C}^3 \oplus {{\A}_{_{1,1}}}.i\big)$, respectively.\\\\
$\bullet \bullet$~The isomorphism of Manin supertriples between $\big({\cal G} , {\cal {I}}_{_{(2|2)}}\big)$ and
$\big({\cal G} , {({\C}^3 + {\A})_{_{k=4}}}^{\hspace{-5mm}\epsilon=-1}\big)$ is given by the following transformation
\begin{eqnarray}\label{A.2}
X'_{_1} & = & \frac{1}{ab} X_{_1},~~~~~~~~~~~~~~~~~~~~ X'_{_5} = -\frac{1}{b} X_{_5}, \nonumber\\
X'_{_2} & = & a^{^2}  X_{_2},~~~~~~~~~~~~~~~~~~~~~X'_{_6} = -c X_{_5}-a  X_{_6},\nonumber\\
X'_{_3} & = & a^{^2}  X_{_2}+ ab X_{_3},~~~~~~~~~~~X'_{_7} = 2a X_{_6}-b X_{_7} +\frac{b c }{a}  X_{_8},\nonumber\\
X'_{_4} & = & -\frac{1}{ab} X_{_1}+\frac{1}{a^{^2}} X_{_4},~~~~~~~X'_{_8} = \frac{2}{b}  X_{_5}-\frac{1}{a} X_{_8},
\end{eqnarray}
where $(X_{_1},\cdots, X_{_8})$ and $(X'_{_1},\cdots, X'_{_8})$ are generators of the Manin supertriple $\big({\cal G} , {\cal {I}}_{_{(2|2)}}\big)$ and\\
$\big({\cal G} , {({\C}^3 +\\ {\A})_{_{k=4}}}^{\hspace{-5mm}\epsilon=-1}\big)$, respectively.
Similarly, one can find an isomorphism transformation between the DSDs
$\big({\cal G} , {\C}^3 \oplus {{\A}_{_{1,1}}}.i\big)$ and  $\big({\cal G} , {({\C}^3 + {\A})_{_{k=4}}}^{\hspace{-5mm}\epsilon=-1}\big)$
that preserves also equations \eqref{2.6} and \eqref{2.7}.

\subsection{Structure of $DSD 4:\big({\cal G} , {{\C}_{_{p=-1}}}^{\hspace{-6mm}1,\epsilon=1} \oplus {{\A}}\big)\cong
\big({\cal G} , {{\C}_{_{p=-1}}}^{\hspace{-6mm}1,\epsilon=-1} \oplus {{\A}}\big)$}

The isomorphism of Manin supertriples between $\big({\cal G} , {{\C}_{_{p=-1}}}^{\hspace{-6mm}1,\epsilon=1} \oplus {{\A}}\big)$ and
$\big({\cal G} , {{\C}_{_{p=-1}}}^{\hspace{-6mm}1,\epsilon=-1} \oplus {{\A}}\big)$ is given by the following transformation
\begin{eqnarray}\label{A.3}
X'_{_1} & = & \frac{1}{ab} (X_{_1} +d X_{_2}),~~~~~~~~~~~~~~~~~~~ X'_{_5} = -\frac{1}{b} X_{_5}, \nonumber\\
X'_{_2} & = &- X_{_2},~~~~~~~~~~~~~~~~~~~~~~~~~~~~~~~~X'_{_6} = -a(c X_{_5}+  X_{_6})+\frac{a^{^2}+1}{2a} X_{_8},\nonumber\\
X'_{_3} & = & a b  (c X_{_2}+  X_{_3}),~~~~~~~~~~~~~~~~~~~~X'_{_7} = b (-X_{_7} + c  X_{_8}),\nonumber\\
X'_{_4} & = & c X_{_1}+cd X_{_2} + d X_{_3}- X_{_4},~~~~~~X'_{_8} = -\frac{1}{a} X_{_8},
\end{eqnarray}
where $(X_{_1},\cdots, X_{_8})$ denotes the basis in the  Manin supertriple
$\big({\cal G} , {{\C}_{_{p=-1}}}^{\hspace{-6mm}1,\epsilon=1} \oplus {{\A}}\big)$, and
$(X'_{_1},\cdots, X'_{_8})$ is the basis in the  Manin supertriple
$\big({\cal G} , {{\C}_{_{p=-1}}}^{\hspace{-6mm}1,\epsilon=-1} \oplus {{\A}}\big)$.

\subsection{Structure of $DSD 5:\big({\cal G} , {{\C}_{_{p=1}}}^{\hspace{-4mm}2,\epsilon=1} \oplus {{\A}_{_{1,1}}}\big)\cong
\big({\cal G} , {{\C}_{_{p=1}}}^{\hspace{-4mm}2,\epsilon=-1} \oplus {{\A}_{_{1,1}}}\big)\\~~~~~~~~~~~~~~~~~~~~~~~~~~~\cong
\big({\cal G} , {{\C}}^{4,\epsilon=1} \oplus {{\A}_{_{1,1}}}\big)\cong
\big({\cal G} , {{\C}}^{4,\epsilon=-1} \oplus {{\A}_{_{1,1}}}\big)$}

The isomorphism transformation of Manin supertriples:\\
$~\bullet \{X_{_A}\} \in \big({\cal G} , {{\C}_{_{p=1}}}^{\hspace{-4mm}2,\epsilon=1} \oplus {{\A}_{_{1,1}}}\big) \longrightarrow
\{X'_{_A}\} \in  \big({\cal G} , {{\C}_{_{p=1}}}^{\hspace{-4mm}2,\epsilon=-1} \oplus {{\A}_{_{1,1}}}\big)$
\begin{eqnarray}\label{A.4}
X'_{_1} & = & -\frac{1}{2ab} X_{_1},~~~~~~~~~~~~~~~~~~~~ X'_{_5} = -\frac{1}{2 b} X_{_8}, \nonumber\\
X'_{_2} & = & X_{_2},~~~~~~~~~~~~~~~~~~~~~~~~~~~X'_{_6} = -\frac{1}{2a} X_{_5}-a  X_{_7}-c X_{_8},\nonumber\\
X'_{_3} & = & -2 a b  X_{_3},~~~~~~~~~~~~~~~~~~~~X'_{_7} = -\frac{2bc}{a} X_{_5} + 2bX_{_6} -b  X_{_8},\nonumber\\
X'_{_4} & = &  X_{_4},~~~~~~~~~~~~~~~~~~~~~~~~~~~X'_{_8} = \frac{1}{a} X_{_5}.
\end{eqnarray}
\\
$~\bullet \bullet \{X_{_A}\} \in \big({\cal G} , {{\C}_{_{p=1}}}^{\hspace{-4mm}2,\epsilon_{_1}} \oplus {{\A}_{_{1,1}}}\big) \longrightarrow
\{X'_{_A}\} \in  \big({\cal G} , {{\C}}^{4,\epsilon_{_2}} \oplus {{\A}_{_{1,1}}}\big)$
\begin{eqnarray}\label{A.5}
X'_{_1} & = & -\frac{1}{2 \epsilon_{_1} ab} X_{_1},~~~~~~~~~~~~~~~~~~~~~~~~~~~ X'_{_5} = -\frac{1}{2 \epsilon_{_1} b} X_{_8}, \nonumber\\
X'_{_2} & = &- \epsilon_{_1} \epsilon_{_2} X_{_2},~~~~~~~~~~~~~~~~~~~~~~~~~~~~~X'_{_6} = \frac{\epsilon_{_2}}{2a} X_{_5}-a  X_{_7},\nonumber\\
X'_{_3} & = & -\epsilon_{_1} \epsilon_{_2}X_{_2} -2 \epsilon_{_1} a b  X_{_3},~~~~~~~~~~~~~~~X'_{_7} =  2 \epsilon_{_1} bX_{_6} -b  X_{_8},\nonumber\\
X'_{_4} & = & \frac{1}{2\epsilon_{_1} ab} X_{_1}-\frac{1}{\epsilon_{_1} \epsilon_{_2}} X_{_4},~~~~~~~~~~~~~~~~X'_{_8} = \frac{1}{a} X_{_5},
\end{eqnarray}
where $\epsilon_{_1}, \epsilon_{_2} =\pm 1$.

$~\bullet \bullet \bullet \{X_{_A}\} \in  \big({\cal G} , {{\C}}^{4,\epsilon=1} \oplus {{\A}_{_{1,1}}}\big) \longrightarrow
\{X'_{_A}\} \in  \big({\cal G} , {{\C}}^{4,\epsilon=-1} \oplus {{\A}_{_{1,1}}}\big)$
\begin{eqnarray}\label{A.6}
X'_{_1} & = & \frac{1}{ab} X_{_1},~~~~~~~~~~~~~~~~~~~~~~~~~~~~~~ X'_{_5} = -\frac{1}{b} X_{_5}, \nonumber\\
X'_{_2} & = &-  X_{_2},~~~~~~~~~~~~~~~~~~~~~~~~~~~~~~~X'_{_6} = -a X_{_6} + \frac{1+a^2}{2a} X_{_8},\nonumber\\
X'_{_3} & = & -(1+ab) X_{_2} + a b  X_{_3},~~~~~~~~~~X'_{_7} = -b  X_{_7},\nonumber\\
X'_{_4} & = & -\frac{1+ab}{ab} X_{_1}-X_{_4},~~~~~~~~~~~~~~~X'_{_8} =- \frac{1}{a} X_{_8}.
\end{eqnarray}

\subsection{Structure of ${DSD 22}^{^p}:\big({\cal G} , { {\D}_{_{p,p-1}}}^{\hspace{-6mm}1,\epsilon=1}\big)\cong
\big({\cal G} , { {\D}_{_{1-p,-p}}}^{\hspace{-8mm}1,\epsilon=1}\big)\cong
\big({\cal G} , { {\D}_{_{p,p-1}}}^{\hspace{-6mm}1,\epsilon=-1}\big),~~~p\neq 0,1$}
The isomorphism transformation of Manin supertriples:
\\
$~\bullet \{X_{_A}\} \in \big({\cal G} , { {\D}_{_{p,p-1}}}^{\hspace{-6mm}1,\epsilon=1}\big) \longrightarrow
\{X'_{_A}\} \in  \big({\cal G} , { {\D}_{_{p',p'-1}}}^{\hspace{-7mm}1,\epsilon=1}\big)$
\begin{eqnarray}\label{A.7}
X'_{_1} & = &a X_{_1} -ac X_{_2},~~~~~~~~~~~~~~~~~~~~ X'_{_5} = \frac{1}{2 b p'} X_{_8}, \nonumber\\
X'_{_2} & = & X_{_2},~~~~~~~~~~~~~~~~~~~~~~~~~~~~~~~~X'_{_6} = \frac{ab p'}{p' -1}  X_{_5} - \frac{1}{ 2ab p'} X_{_7},\nonumber\\
X'_{_3} & = & \frac{1}{a}  X_{_3},~~~~~~~~~~~~~~~~~~~~~~~~~~~~~~X'_{_7} = - 2b p' X_{_6} -b  X_{_8},\nonumber\\
X'_{_4} & = &  c X_{_3} +X_{_4},~~~~~~~~~~~~~~~~~~~~~~~~X'_{_8} = 2ab p' X_{_5},
\end{eqnarray}
where $p' = 1-p$.
\\
$~\bullet \bullet \{X_{_A}\} \in \big({\cal G} , { {\D}_{_{p,p-1}}}^{\hspace{-6mm}1,\epsilon=1}\big) \longrightarrow
\{X'_{_A}\} \in  \big({\cal G} , { {\D}_{_{p,p-1}}}^{\hspace{-6mm}1,\epsilon=-1}\big)$
{\small \begin{eqnarray}\label{A.8}
X'_{_1} & = &a X_{_1},~~~~~~~~~~~~~~~~~~~~~~~ X'_{_5} =- \frac{1}{b} X_{_5}, \nonumber\\
X'_{_2} & = & -X_{_2},~~~~~~~~~~~~~~~~~~~~~~X'_{_6} =-\frac{1}{ab}  X_{_6} - c X_{_8},\nonumber\\
X'_{_3} & = & \frac{1}{a}  X_{_3},~~~~~~~~~~~~~~~~~~~~~~X'_{_7} =  -b  X_{_7},\nonumber\\
X'_{_4} & = &  -X_{_4},~~~~~~~~~~~~~~~~~~~~~~X'_{_8} = -ab X_{_8},
\end{eqnarray}}
where $c=\frac{1+a^2 b^2}{2ab(1-p)}$.
\subsection{Structure of ${DSD 23}^{^k}: \big({\cal G} , {(2{\A}_{_{1,1}} + 2{\A})^{^{0}}}.i\big)\cong
\big({\cal G} , {({ {\C}}^3 + {\A})_{_{k-4}}}^{\hspace{-4mm}\epsilon=1}\big)\\~~~~~~~~~~~~~~~~~~~~~~~~~~~~~\cong
\big({\cal G} , {({ {\C}}^3 + {\A})_{_{k}}}^{\hspace{-2mm}\epsilon=-1}\big),~ k>4$}

The isomorphism transformation of Manin supertriples:
\\
{\small $~\bullet \{X_{_A}\} \in \big({\cal G} , {(2{\A}_{_{1,1}} + 2{\A})^{^{0}}}.i\big) \longrightarrow
\{X'_{_A}\} \in  \big({\cal G} , {({ {\C}}^3 + {\A})_{_{k}}}^{\hspace{-2mm}\epsilon=-1}\big)$}
{\small\begin{eqnarray}\label{A.9}
X'_{_1} & = & \frac{1}{ab} X_{_1},~~~~~~~~~~~~~~~~~~~~~~~~~~~~~~~~ X'_{_5} = -\frac{1}{b} X_{_5}, \nonumber\\
X'_{_2} & = & a^2 X_{_2},~~~~~~~~~~~~~~~~~~~~~~~~~~~~~~~~~X'_{_6} = -a X_{_6},\nonumber\\
X'_{_3} & = & \frac{a c^2}{b-2c} X_{_2} + ab X_{_3},~~~~~~~~~~~~~~~~~X'_{_7} = - c X_{_6} -b  X_{_7},\nonumber\\
X'_{_4} & = &  \frac{c^2}{a^2 b(2c-b)} X_{_1} +\frac{1}{a^2} X_{_4},~~~~~~~~~~X'_{_8} = -\frac{c}{ab}  X_{_5}-\frac{1}{a}  X_{_8},
\end{eqnarray}}
where $c(c-b)=a(b-2c)$ and $k=\frac{b-2c}{a} >4$.
\\
$~\bullet \bullet \{X_{_A}\} \in \big({\cal G} , {({ {\C}}^3 + {\A})_{_{k-4}}}^{\hspace{-4mm}\epsilon=1}\big) \longrightarrow
\{X'_{_A}\} \in \big({\cal G} , {({ {\C}}^3 + {\A})_{_{k}}}^{\hspace{-2mm}\epsilon=-1}\big)$
\begin{eqnarray}\label{A.10}
X'_{_1} & = &\frac{1}{a^2} X_{_1},~~~~~~~~~~~~~~~~~~~~~~ X'_{_5} =- \frac{1}{a} X_{_5}, \nonumber\\
X'_{_2} & = & a^2 X_{_2},~~~~~~~~~~~~~~~~~~~~~~~X'_{_6} = 2b X_{_5} -a  X_{_6} - b X_{_8},\nonumber\\
X'_{_3} & = & a^2 X_{_3},~~~~~~~~~~~~~~~~~~~~~~~X'_{_7} = 2a  X_{_6} -a X_{_7},\nonumber\\
X'_{_4} & = &  \frac{1}{a^2} X_{_4},~~~~~~~~~~~~~~~~~~~~~~X'_{_8} = \frac{2}{a} X_{_5} -\frac{1}{a} X_{_8}.
\end{eqnarray}

\subsection{Structure of ${DSD 24}^{^k}:\big({\cal G} , {({{\C}}^3 + {\A})_{_{k}}}^{\hspace{-2mm}\epsilon=-1}\big)\cong
\big({\cal G} , {{({\C}}^3 + {\A})_{_{4-k}}}^{\hspace{-4mm}\epsilon=-1}\big),~0<k<4$}

The isomorphism of Manin supertriples between $\big({\cal G} , {({{\C}}^3 + {\A})_{_{k}}}^{\hspace{-2mm}\epsilon=-1}\big)$ and
$\big({\cal G} , {{({\C}}^3 + {\A})_{_{4-k}}}^{\hspace{-4mm}\epsilon=-1}\big)$ is given by the following transformation
\begin{eqnarray}\label{A.11}
X'_{_1} & = & -\frac{1}{a^2} X_{_1},~~~~~~~~~~~~~~~~~~~~~ X'_{_5} = -\frac{1}{a} X_{_5}, \nonumber\\
X'_{_2} & = & a^2 X_{_2},~~~~~~~~~~~~~~~~~~~~~~~~X'_{_6} = -2b X_{_5}+ a X_{_6}-b X_{_8},\nonumber\\
X'_{_3} & = & -a^2 X_{_3},~~~~~~~~~~~~~~~~~~~~~~X'_{_7} = -2a X_{_6} -a X_{_7},\nonumber\\
X'_{_4} & = & \frac{1}{a^2} X_{_4},~~~~~~~~~~~~~~~~~~~~~~~X'_{_8} = \frac{2}{a} X_{_5} +\frac{1}{a} X_{_8},
\end{eqnarray}
where $(X_{_1},\cdots, X_{_8})$ denotes the basis in the  Manin supertriple
$\big({\cal G} , {({{\C}}^3 + {\A})_{_{k}}}^{\hspace{-2mm}\epsilon=-1}\big)$, and
$(X'_{_1},\cdots, X'_{_8})$ is the basis in the  Manin supertriple
$\big({\cal G} , {{({\C}}^3 + {\A})_{_{4-k}}}^{\hspace{-4mm}\epsilon=-1}\big)$.


\end{document}